\documentclass{article}
\usepackage{arxiv}
\usepackage{graphicx}
\usepackage{mathptmx}
\usepackage[subrefformat=parens,labelformat=parens]{subcaption}
\usepackage{authblk}
\usepackage{comment}
\usepackage[titletoc]{appendix}

\usepackage{amsfonts}
\usepackage{algorithm}
\usepackage{amsbsy}
\usepackage{amsthm}
\usepackage{amsmath}

\usepackage{algpseudocode}
\usepackage{bbm}
\usepackage{wrapfig}
\usepackage{booktabs}
\usepackage{multirow}
\usepackage{wrapfig,lipsum,booktabs}
\usepackage{hyperref}
\allowdisplaybreaks
\usepackage{colortbl}
\usepackage{seqsplit}
\usepackage{xcolor}
\definecolor{KWgreen}{RGB}{112,173,71} 
\definecolor{KWblue}{RGB}{0,112,192} 
\definecolor{KWred}{RGB}{192,0,0} 
\definecolor{KWpurple}{RGB}{112,48,160} 
\usepackage{float}

\usepackage[authoryear,longnamesfirst]{natbib}
\usepackage{booktabs}
\usepackage{tabularx}
\usepackage{threeparttablex}
\usepackage{makecell}
\usepackage{float}  
\usepackage{placeins}
\usepackage{amsmath}
\usepackage{longtable}
\title{Intracellular Measurement‑Informed Multiscale Modeling for Scalable iPSC Manufacturing
}

\begin{document}

\author[1]{Fuqiang Cheng}
\author[3]{Zahra Foroozan Jahromi}
\author[1]{Keqi Wang}
\author[2]{Thomas C. Caldwell}
\author[4]{Grace Cai}
\author[1]{Keilung Choy}
\author[1]{Jared Auclair}
\author[4]{Jeffrey L. Campbell}
\author[4]{Youbo Zhao}
\author[5]{Jane Ring}
\author[3]{Seongkyu Yoon}
\author[2]{Sarah W. Harcum}
\author[1]{Wei Xie\thanks{Corresponding author: w.xie@northeastern.edu}}

% Include full affiliation details for all authors
\affil[1]{Department of Mechanical and Industrial Engineering, Northeastern University, Boston, MA 02115, USA}

\affil[2]{Department of Bioengineering, Clemson University, Clemson, SC, USA}

\affil[3]{Department of Chemical Engineering, University of Massachusetts Lowell, Lowell, MA, USA}

\affil[4]{Physical Sciences Inc. (PSI), Andover, MA, USA}

\affil[5]{MilliporeSigma, Bedford, MA, USA}
\maketitle

\begin{abstract}
Scalable manufacturing of human induced pluripotent stem cells (iPSCs) is essential for industrial‑scale production of cell therapies and regenerative medicines. However, the 3D aggregate cultures used in manufacturing exhibit substantial spatial and metabolic heterogeneity compared with the relatively homogeneous monolayer systems used in laboratory studies, complicating mechanistic understanding and predictive metabolic modeling across culture scales.
To address this challenge, we developed a modular multiscale mechanistic foundation model that links molecular, cellular, and macroscopic processes while accounting for spatial and metabolic heterogeneity. The framework integrates extracellular culture dynamics, intracellular metabolic fluxes, and cellular redox states by extending a previously established monolayer kinetic network and coupling it with a biological systems-of-systems (Bio‑SoS) multiscale model for aggregate cultures, incorporating explicit redox interactions. Systematic monolayer and aggregate experiments—including 
multiple isotopic tracers, extracellular metabolite profiling, and two‑photon optical redox imaging—were used to improve and validate the model. This integrated framework unifies heterogeneous datasets across culture configurations and enables mechanistic interpretation of metabolic and redox responses across heterogeneous culture scales, providing a quantitative foundation for scalable iPSC biomanufacturing.
\end{abstract}


\section{Introduction}

Human induced pluripotent stem cells (iPSCs) serve as a versatile platform for disease modeling, regenerative medicines, and cell-based therapeutics due to its capacity for indefinite self-renewal and multilineage differentiation
\citep{takahashi2007induction,yu2007induced}. As iPSC‑derived products progress toward clinical translation and commercial‑scale manufacturing \cite{nogueira2019strategies, yang2026multiscale}, there is a growing need for robust, scalable culture strategies that preserve metabolic stability, redox homeostasis, and phenotypic consistency
\citep{chen2011chemically,serra2012process}. Suspension-based and aggregate-forming culture systems have emerged as particularly promising for large‑scale production, as they integrate efficiently with stirred‑tank bioreactors and controlled manufacturing environments \citep{olmer2012suspension,serra2012process}.

{Yet, three‑dimensional (3D) aggregate cultures introduce significant mass-transfer challenges because oxygen and nutrients must diffuse from the aggregate surface to the core. Experimental studies have demonstrated the development of oxygen limitations in multicellular spheroids and stem-cell aggregates as aggregate size increases \citep{mueller1984method,amit2010suspension,gupta2016optimization,wu2014oxygen}. In addition, iPSC growth and metabolism have been shown to be highly sensitive to dissolved oxygen availability \citep{amit2010suspension}, while different aggregate size distributions have been associated with distinct metabolite consumption and production rates in stirred-tank iPSC cultures \citep{schwedhelm2019automated}. Together, these observations suggest the presence of intra-aggregate gradients in oxygen, nutrients, and metabolic activity, resulting in heterogeneous intracellular states across the aggregate radius. Such gradients can significantly alter glycolytic flux, mitochondrial activity, and redox balance, thereby complicating quantitative metabolic characterization and limiting the accuracy of predictive process control in scalable iPSC culture systems \citep{zhang2012metabolic,Folmes2011MetabolicPlasticity}.}

%connection of redox with metabolism [add agg]
Cellular metabolism plays a central role in regulating iPSC growth and maintaining pluripotency \cite{shyh2013stem}. Compared to differentiated cells, pluripotent stem cells rely heavily on glycolysis, exhibiting elevated glycolytic flux, high lactate production, and reduced dependence on oxidative phosphorylation \citep{Varum2011Energy,Folmes2011MetabolicPlasticity}. Central carbon metabolism through glycolysis and the tricarboxylic acid (TCA) cycle is tightly coupled to intracellular redox reactions involving NAD$^+$/NADH, NADP$^+$/NADPH, and FAD/FADH$_2$, thereby linking extracellular oxygen and nutrient availability to cellular energy production, biosynthetic pathways, and signaling networks.

\begin{sloppypar}
In 2D static monolayer cultures, cells experience relatively uniform exposure to oxygen and nutrients, enabling the metabolic states to be interpreted under conditions that are effectively homogeneous. In contrast, 3D aggregate cultures impose diffusion limitations that create spatial gradients in oxygen tension and substrate availability. These gradients can drive pronounced shifts in glycolytic activity, mitochondrial flux, and redox balance across the aggregate radius, even when bulk extracellular measurements appear comparable to those of 2D systems.

Because intracellular redox state is an emergent property of metabolic flux distributions and is particularly sensitive to local transport constraints, predictive modeling of scalable iPSC cultures must integrate metabolic regulation with substrate and waste transport and redox dynamics in a unified framework. Only by treating these processes as a tightly interconnected system can model predictions accurately capture the spatial and metabolic heterogeneity inherent to 3D culture environments.

\end{sloppypar}

%rather than treating them as independent phenomena.

Advances in non‑invasive optical sensing have further expanded the ability to integrate metabolic and redox measurements into iPSC culture analysis. Autofluorescence imaging of NAD(P)H and FAD provides label‑free assessment of intracellular redox state in living cells \citep{Skala2007InVivoNADH, kolenc2019evaluating, blacker2016investigating}.  In particular, two‑photon excitation (TPE) sensing offers real‑time quantification of intracellular metabolites and the NAD(P)H/FAD redox ratio, enabling dynamic monitoring of metabolic state changes within complex culture environments.  When implemented in bioreactors, TPE benefits both from its intrinsic optical selectivity and from the high flow dynamics of agitated cultures, producing cytometry‑like, single‑cell–level metabolic readouts in situ—capabilities that are difficult to achieve in large‑scale suspension systems. Consequently, TPE sensing is well suited for aggregate‑based iPSC cultures, where destructive intracellular sampling is impractical and spatial heterogeneity complicates the interpretation of conventional biochemical assays.

Quantitative metabolic modeling has been widely applied in mammalian cell cultures to support rational biomanufacturing process development and improve understanding of cellular metabolism \cite{nolan2011dynamic, ghorbaniaghdam2014analyzing}. However, translating these approaches to iPSC systems presents additional challenges.
Unlike many industrial mammalian production cell lines, which operate under relatively stable productivity-driven objectives, iPSCs exhibit metabolism tightly linked to pluripotency maintenance and redox-sensitive regulatory networks. Furthermore, scalable culture formats such as suspension aggregates introduce diffusion-limited transport of oxygen and nutrients, leading to spatially heterogeneous microenvironments and dynamically varying intracellular metabolic states. These biological and process-specific characteristics complicate direct interpretation of extracellular measurements and highlight the need for mechanistic modeling frameworks tailored to iPSC cultures. Such models must link extracellular culture dynamics with intracellular metabolic flux regulation while accounting for heterogeneity arising from aggregate-scale transport phenomena.

\begin{sloppypar}
Recent efforts have begun to advance quantitative iPSC culture modeling frameworks. Wang et al. developed a kinetic metabolic regulatory network model that integrates multiple isotopic tracers to improve intracellular flux analysis in 2D monolayer iPSC cultures, providing a structured representation of glycolysis, TCA cycle activity, and associated regulatory interactions under relatively homogeneous conditions \citep{Wang2024}. Recognizing that individual cells operate as complex systems and that  {cell–cell and cell–extracellular matrix interactions generate heterogeneous microenvironments},  Zheng et al. introduced a biological systems‑of‑systems (Bio-SoS) modeling strategy and a modular, multiscale mechanistic foundational model capable of assembling both 2D monolayer and 3D aggregate cultures \citep{zheng2024stochastic}. 
Their framework captures multiscale variability by modeling spatial and metabolic heterogeneity and population-level dynamics in iPSC aggregates—key determinants of quality consistency and cross-scale integration in manufacturing.
\end{sloppypar}

%Together, these studies provide important methodological foundations for linking extracellular measurements with intracellular metabolic states. 

However, existing iPSC culture modeling frameworks either prioritize flux fitting under near‑homogeneous monolayer conditions or center on aggregate‑scale heterogeneity. To our knowledge, no previous iPSC modeling framework has been systematically calibrated and evaluated using complementary experimental observations spanning both monolayer and aggregate cultures. In addition, they do not explicitly account for intracellular redox dynamics. Consequently, quantitative relationships linking nutrient perturbations, intracellular flux redistribution, redox balance, and aggregate growth dynamics remain incompletely characterized.

Building on the kinetic metabolic regulatory network model for monolayer iPSC cultures \citep{Wang2024} and the multiscale Bio-SoS framework for 3D aggregate cultures \citep{zheng2024stochastic}, this study further extends and experimentally validates a unified multiscale mechanistic foundation model. First, the intracellular metabolic network is expanded to explicitly represent redox coupling, establishing a direct connection between metabolic flux distributions and cellular redox dynamics through NAD\textsuperscript{+}/NADH- and FAD/FADH$_2$-associated reactions.  {Second, the multiscale framework is further refined through integration of reaction--diffusion transport and population balance modeling to connect extracellular nutrient and oxygen gradients with heterogeneous intracellular metabolic states, while refined diffusion parameters and explicit oxygen transport representation improve characterization of intra-aggregate microenvironments.} Third, systematic isotopic experiments performed under static and pyruvate-supplemented culture conditions uncover new metabolic mechanisms that advance understanding of TCA cycle activity and metabolism--redox interactions.  {Finally, the extended multiscale foundation model is calibrated and evaluated using comprehensive experimental datasets spanning both monolayer and aggregate cultures.} Together, these advances provide a quantitative framework for improved characterization of iPSC metabolic behavior and support the acceleration of scalable and interoperable iPSC manufacturing process development.

\section{Materials and Methods}

\subsection{Cell Line}
K3 induced pluripotent stem cells (iPSCs) were donated by Dr. Stephen A. Duncan at the Medical University of South Carolina. The K3 iPSCs were generated by transfection of human foreskin fibroblasts, as previously described \citep{si2010generation}. 

\subsection{Pre-Culture Conditions}
K3 iPSCs were cultured on non–tissue culture–treated 6-well plates (Corning Inc., Corning, NY) coated with 10~$\mu$g/mL Vitronectin XF (Catalog \#07180; StemCell) according to the manufacturer’s instructions. Cultures were maintained at 37$^{\circ}$C in a humidified incubator with 5\% CO$_2$ using Essential~8 Flex (E8 Flex) medium (Thermo Fisher Scientific, Waltham, MA). The working volume was 2~mL per well. Cells were passaged every 3~days at split ratios of 1:6 or 1:12 depending on confluency.

For passaging, cells were dissociated using Gibco StemPro Accutase Cell Dissociation Reagent (Thermo Fisher Scientific) for 2~min at 37$^{\circ}$C, followed by washing with phosphate-buffered saline (PBS) without calcium or magnesium (Corning Inc.). Detached cells were centrifuged at 500 $\times ~g$ for 5~min and resuspended in fresh E8 Flex supplemented with 10~$\mu$M Y-27632, a rho-associated coiled-coil containing protein kinase inhibitor (ROCKi; STEMCELL Technologies, Vancouver, Canada). After 24~h, the medium was replaced with E8 Flex without ROCKi.

\subsection{Static Pyruvate Cultures: Growth and Parallel Labeling}
For growth and isotope labeling studies, K3 iPSCs were seeded at a density of $1 \times 10^{4}$~cells/cm$^{2}$ in E8 Flex supplemented with 10~$\mu$M ROCK inhibitor to promote attachment and reduce apoptosis. After 24~h, a complete media exchange was performed.
One of four defined growth media (2 glucose levels × 2 lactate levels) was added to parallel 6-well plates and T75 flasks. An overview of the experimental design, including initial concentrations of glucose, glutamine, lactate, and pyruvate, is provided in Table~\ref{tab:exp_conditions}. E8 Flex was prepared without glucose to enable controlled adjustment of glucose levels. For isotope‑labeling experiments, 4~mM [U-$^{13}$C$_3$] pyruvate (99\% enrichment; Cambridge Isotope Laboratories, Tewksbury, MA) was used.

\subsection{Aggregate Cultures}
Suspension aggregate cultures were established following previously reported protocols \citep{CuestaGomez2023}. Briefly, static culture cells were dissociated using Accutase, and the reaction was neutralized with pre-warmed medium. Cells were centrifuged at 200 $\times ~g$ for 4~min at 25$^{\circ}$C, and the supernatant was discarded. The cell pellet was gently resuspended in approximately 10~mL of pre-warmed medium to generate a single-cell suspension. Cells were counted and resuspended in E8 Flex supplemented with 10~$\mu$M ROCKi prior to aggregate formation.

\subsection{Analytical Methods}

\subsubsection{Cell Concentration and Metabolite Analysis}
Cell concentration and extracellular glucose, lactate, pyruvate, ammonia, and amino acid concentrations were measured for static pyruvate and aggregate cultures using methods summarized in Table~\ref{tab:exp_conditions} and in accordance with standard manufacturing protocols. For aggregate cultures, a 1~mL sample was collected and aggregates were pelleted by centrifugation at 500~$\times~ g$ for 5~min. The supernatant was retained for metabolite and amino acid analysis. The cell pellet was subsequently dissociated using Gibco StemPro Accutase and resuspended in 1~mL of fresh medium for cell counting and viability assessment.

\subsubsection{Aggregate Size Distribution}
Aggregate size distributions were measured daily using one shaker flask per condition. To assess aggregate morphology and size, 1–2~mL of culture was transferred into 12-well plates or petri dishes and allowed to settle prior to imaging. Brightfield images were acquired using an Olympus microscope equipped with a digital camera (cellSens Standard 3.2). Aggregate diameters and counts were quantified using ImageJ (NIH). Images were background-subtracted, converted to binary format, and filtered by circularity to exclude debris and non-aggregate objects. Size distributions were calculated from more than 50 aggregates per sample using the \textit{Analyze Particles} function.

\subsubsection{Enzymatic Redox Assay}
Intracellular concentrations of NADH and NAD$^{+}$ were quantified using the EnzyChrom NAD$^{+}$/NADH Assay Kit (BioAssay Systems, Hayward, CA) following the manufacturer’s protocol. One shaker flask was harvested per time point, and 1~mL of aggregates was enzymatically dissociated using Accutase as described above. Fluorescence measurements were obtained using a BioTek Synergy H1 microplate reader, with data acquisition and analysis performed using Gen5 software (BioTek Instruments, Winooski, VT). The redox ratio was calculated as %\text{Redox ratio} =
$\frac{\mathrm{NAD}^{+}}{\mathrm{NAD}^{+} + \mathrm{NADH}}$
using normalized fluorescence intensities.

\subsubsection{TPE Redox Measurement}
For optical redox measurements using the two-photon excitation (TPE) system (Prototype-002, PSI), the contents of the shaker flask were transferred to a beaker and placed on a pre-warmed stirring platform. Autofluorescence signals from NAD(P)H and FAD were recorded  in real time at the single cell level. The optical redox ratio was calculated as $\frac{\mathrm{FAD}}{\mathrm{FAD} + \mathrm{NAD(P)H}}$, and redox estimates were obtained using PSI’s proprietary data analysis software. This redox ratio is widely used as an optical indicator of cellular metabolic state derived from endogenous fluorescence signals \citep{quinn2013quantitative}. It reflects the balance between oxidized flavin cofactors and reduced nicotinamide cofactors and has been shown to correlate with cellular metabolic shifts.

Although the enzymatic assay quantifies the ratio of NAD$^{+}$ to total nicotinamide cofactors, while the optical measurement captures the fluorescence ratio of FAD and NAD(P)H, both metrics represent the balance between oxidized and reduced cofactors associated with cellular metabolic activity \citep{quinn2013quantitative}. Agreement between trends from the enzymatic assay and the TPE‑derived optical redox ratio therefore supports the use of the TPE sensor as a non‑destructive method for monitoring intracellular redox dynamics.

\begin{table*}[htbp]
\centering
\footnotesize
\begin{threeparttable}
\caption{Experimental conditions for the Historic Static \citep{Odenwelder2021}, Static Pyruvate, and Aggregate cultures.}
\label{tab:exp_conditions}

\begin{tabularx}{\textwidth}{l p{3.5cm} X X}
\toprule
\textbf{Experiment} 
& \textbf{Historic Static Cultures} 
& \textbf{Static Pyruvate Cultures} 
& \textbf{Aggregate Cultures} \\
\midrule

\multicolumn{4}{l}{\textbf{Media Composition}} \\

Initial glucose (mM) (LG/HG) 
& 5.6 / 18.3 
& 10.4 / 33.6 
& 17.0 \\

Initial glutamine (mM) 
& 2.7 
& 4.69 
& 1.84 \\

Initial lactate (mM) (LL/HL) 
& 0 / 20 
& 0 / 38 
& 0 \\

Initial pyruvate (mM) 
& 0.41 
& 4.4 
& 0 \\

Initial cell density or equivalent (cells/mL) 
& $3.2 \times 10^{4}$ 
& $4.8 \times 10^{4}$ 
& $2.0 \times 10^{5}$ \\

Initial cell density (cells/cm$^{2}$) 
& $\sim 1 \times 10^{4}$ 
& $\sim 1 \times 10^{4}$ 
& N/A \\

Initial pH 
& N/A 
& N/A 
& 7.5 \\

Final pH 
& N/A 
& N/A 
& 6.5 \\

Labeled substrates 
& \raggedright
\text{[1,2-$^{13}$C$_2$] glucose}\newline
\text{[U-$^{13}$C$_5$] glutamine}\newline
\text{[U-$^{13}$C$_3$] lactate}

& [U-$^{13}$C$_3$] pyruvate
& No \\

Other additions 
& 20 mM NaCl (HGLL, LGLL) 
& 20 mM NaCl (HGLL, LGLL) 
& N/A \\

\midrule
\multicolumn{4}{l}{\textbf{Culture Setup}} \\

Culture method 
& 6-well plates/100-mm Petri dishes 
& 6-well plates/T75 flasks 
& Shake flasks (125 mL) \\

Working volume (mL) 
& 3/30 
& 2/20 
& 22--28 \\

\midrule
\multicolumn{4}{l}{\textbf{Environmental Conditions}} \\

Temperature ($^{\circ}$C) 
& 37 
& 37 
& 37 \\

CO$_2$ concentration (\%) 
& 5 
& 5 
& 5 \\

Agitation rate (rpm) 
& N/A 
& N/A 
& \makecell[l]{70 (D0--D1) \\ 75 (D2--D3)} \\

Shake flask type 
& N/A 
& N/A 
& Baffled, PC \\

\midrule
\multicolumn{4}{l}{\textbf{Process Parameters}} \\

Passage number 
& P3 
& P3 
& P3 \\

Culture duration (days) 
& 2 
& 2 
& 3 \\

% Feeding (day) 
% & N/A 
% & N/A 
% & 3 \\

\midrule
\multicolumn{4}{l}{\textbf{Data Collection}} \\

VCD method 
& Vi-Cell XR (Beckman) 
& Vi-Cell XR (Beckman) 
& Nova Flex 2 \\

VCD replicates 
& 6 
& 2 
& 3 \\

Metabolites 
& Cedex Bio 
& Cedex Bio 
& Nova Flex 2 \\

Metabolite replicates 
& 6 
& 2 
& 3 \\

Amino acid analysis method 
& EZ:faast kit (Phenomenex)
& REBEL (908 Devices) 
& REBEL (908 Devices) \\

AA replicates (Tech/Bio) 
& 1 / 6 
& 2 / 2 
& 2 / 3 \\

Redox kit replicates 
& N/A 
& N/A 
& 1 \\

Redox TPE 
& N/A 
& N/A 
& 2 mL in 50 mL conical tube \\

\bottomrule
\end{tabularx}

\begin{tablenotes}[flushleft]
\footnotesize
\item \textbf{Abbreviations:}
Tech = technical (repeated-measures) replicates;
Bio = biological replicates;
HG = high glucose; LG = low glucose;
HL = high lactate; LL = low lactate;
D = day of culture;
PC = polycarbonate;
P = passage number.
\end{tablenotes}

\end{threeparttable}
\end{table*}

\section{Experiment Results and Discussion}
\subsection{Growth Characteristics of Static Pyruvate and Aggregate Cultures}

 \begin{sloppypar}
To advance the characterization of iPSC metabolism and redox regulation, this study employed isotopically labeled pyruvate across combinations of high and low glucose (HG, LG) and high and low lactate (HL, LL) conditions. This investigation builds upon our previously established iPSC monolayer culture system and comprehensive isotope‑labeling experiments. Odenwelder et al. \citep{Odenwelder2021} evaluated iPSC metabolism under four media formulations using labeled glucose and glutamine to resolve intracellular fluxes. Although labeled lactate was also assessed, the resulting intracellular enrichment was insufficient for inclusion in metabolic flux analysis. Motivated by this limitation, we hypothesized that labeled pyruvate would provide more robust intracellular labeling patterns.

Preliminary experiments were conducted to determine pyruvate tolerance, revealing that concentrations of 20.0~mM and 10.0 mM completely inhibited cell growth. Consequently, 4.0 mM pyruvate was selected for the isotope labeling experiments in  high glucose low lactate (HGLL), high glucose high lactate (HGHL), low glucose high lactate (LGHL), and low glucose low lactate (LGLL) conditions. 
 \end{sloppypar}

\begin{figure*}[htbp]
    \centering
    \includegraphics[width=\textwidth]{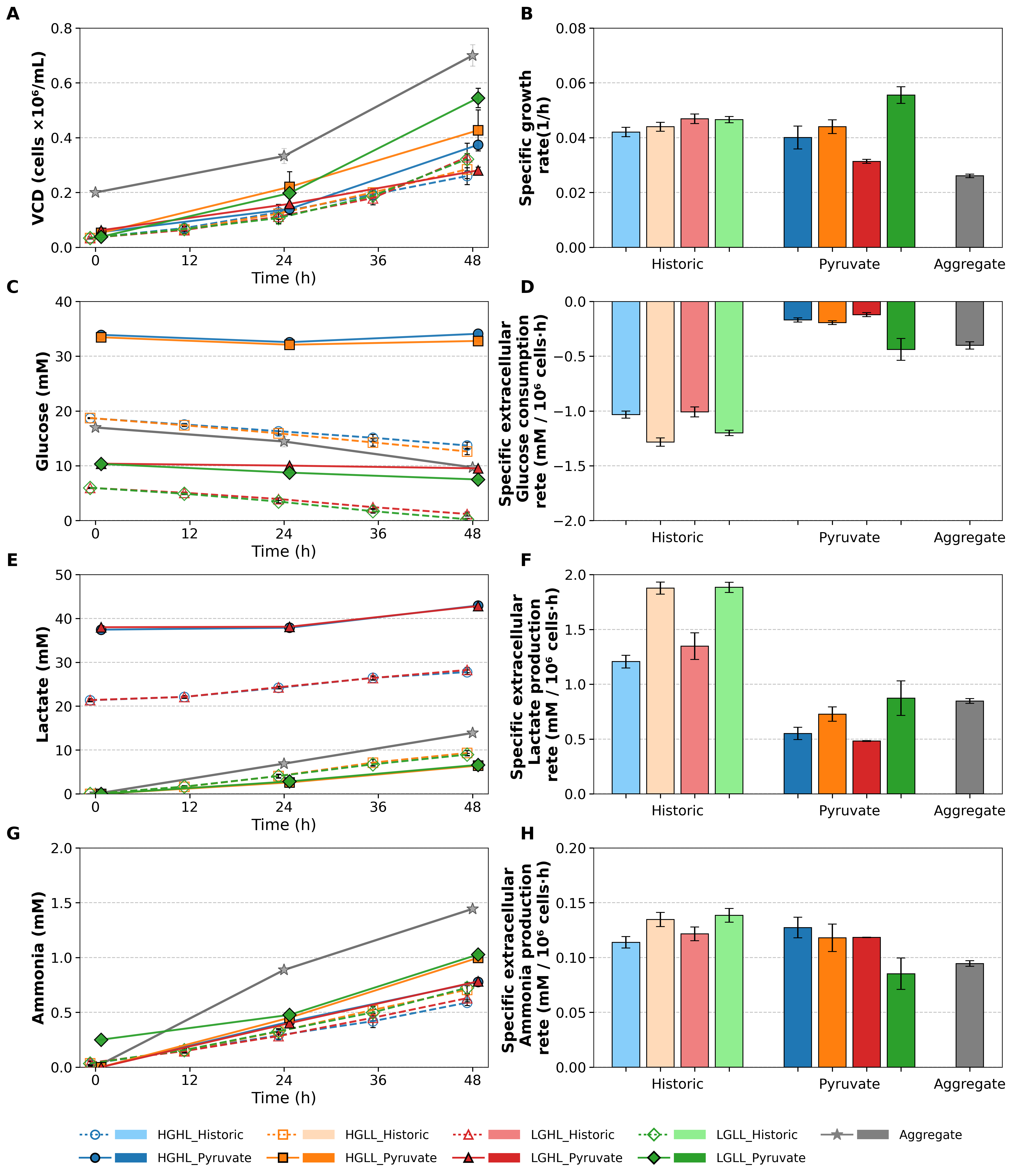}
    \caption{Cell growth, glucose, lactate, and ammonia profiles and rates for
    Historic Static, Static Pyruvate, and Aggregate culture conditions.
    (A) Viable cell density (VCD) profiles.
    (B) Specific growth rates.
    (C) Glucose concentration profiles.
    (D) Specific glucose consumption rates.
    (E) Lactate concentration profiles.
    (F) Specific lactate production rates.
    (G) Ammonia concentration profiles.
    (H) Specific ammonia production rates.
    For clarity, data points corresponding to Historic Static conditions are slightly shifted to the left of each sampling time, whereas Static Pyruvate conditions are slightly shifted to the right to facilitate visual distinction of overlapping measurements. Error bars represent the standard deviation across biological replicates. }

    \label{fig:metabolite_con_rate_1}
\end{figure*}

\begin{figure*}[htbp]
    \centering
    \includegraphics[width=\textwidth]{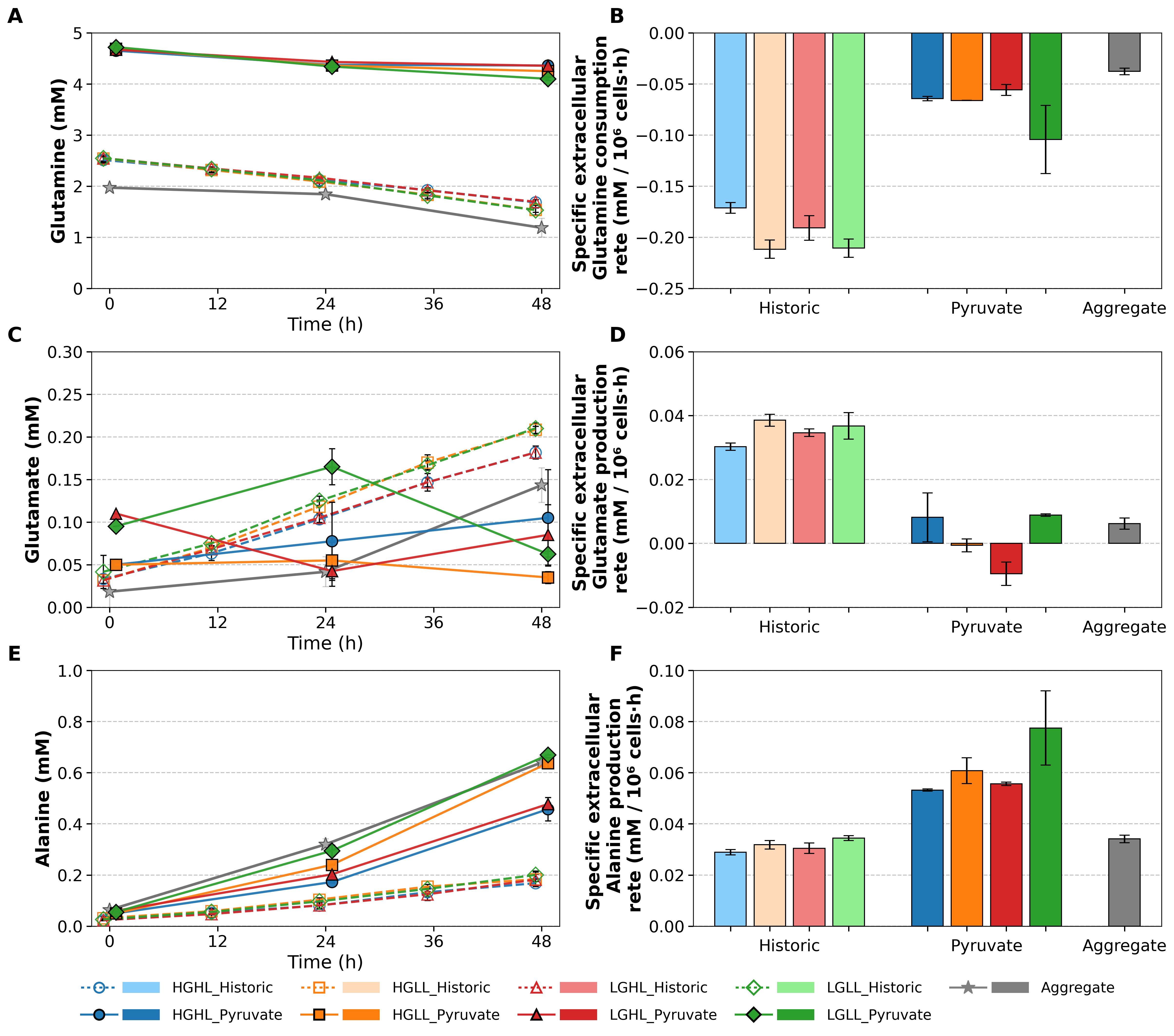}
    \caption{Glutamine, glutamate, and alanine profiles and rates for
    Historic Static, Static Pyruvate, and Aggregate culture conditions.
    (A) Glutamine concentration profiles.
    (B) Specific glutamine consumption rates.
    (C) Glutamate concentration profiles.
    (D) Specific glutamate production rates.
    (E) Alanine concentration profiles.
    (F) Specific alanine production rates.
    Error bars represent the standard deviation across biological replicates. For clarity, data points corresponding to Historic Static conditions are slightly shifted to the left of each sampling time, whereas Static Pyruvate conditions are slightly shifted to the right to facilitate visual distinction of overlapping measurements. Error bars represent the standard deviation across biological replicates. }

    \label{fig:metabolite_con_rate_2}
\end{figure*}

The growth characterization of the Static Pyruvate cultures was compared to the Historic Static cultures \citep{Odenwelder2021} as well as Aggregate cultures in standard E8 Flex media. Figs.~\ref{fig:metabolite_con_rate_1} and \ref{fig:metabolite_con_rate_2} compare the growth and metabolic profiles of these three experimental setups. The initial glucose, glutamine, and pyruvate concentration for Static Pyruvate cultures were significantly higher than that used for the Historic Static cultures and Aggregate cultures, and the high lactate conditions were also significantly higher for the Static Pyruvate cultures. Interestingly, the growth rates for all culture conditions were similar; however, the glucose consumption rate was significantly lower for the Static Pyruvate cultures due to the presence of pyruvate. 

Additionally, glutamine consumption and glutamate production ratio were significantly lower for the Static Pyruvate cultures, which may also be due to pyruvate inhibition \citep{Yang2014}. 
Alanine production was higher for the Static Pyruvate Cultures compared to the Historic Static cultures and Aggregate cultures, likely due to direct alanine synthesis from pyruvate. Several amino acids exhibited reduced consumption rates in Static Pyruvate Cultures compared with Historic Static Cultures. In general, amino acid inhibition mechanisms due to high pyruvate concentration have not been systematically studied in the literature and require further investigation.

Furthermore, lower glucose consumption and lactate production were observed in aggregate cultures, likely due to lower oxygen availability throughout the aggregates \citep{wu2014oxygen,manstein2021high}. In the present study, oxygen transport was incorporated into the reaction-diffusion framework; however, oxygen-specific transport parameters were not independently calibrated because dissolved oxygen measurements within aggregates were unavailable. Glutamine consumption was also lower in aggregate cultures compared with historic static cultures, while glutamate production was reduced and more similar to that observed in static pyruvate cultures. One possible reason is the inhibitory effect of pyruvate on glutamine utilization \citep{Yang2014}. Finally, ammonia production was similar across all three culture conditions, potentially due to counterbalancing effects of reduced glutamine consumption and glutamate production. 

\subsection{Isotopic Analysis of Static Pyruvate Cultures}
%Mass Isotopomer Distribution (MID) Results for the Static Pyruvate Cultures}

To improve isotopic resolution of metabolic fluxes, 4.0 mM [U-\textsuperscript{13}C\textsubscript{3}] 
pyruvate was added to the Static Pyruvate cultures for the labeling experiments. This tracer was incorporated into the E8 Flex medium, yielding a final total pyruvate concentration of 4.4 mM. For consistency, the unlabeled control cultures used for growth and metabolite measurements also received 4.0 mM pyruvate. This additional pyruvate supplement was applied across all four glucose/lactate conditions (HGLL, HGHL, LGHL, and LGLL).

\begin{figure*}[t]
    \centering
    \includegraphics[width=0.66\textwidth]{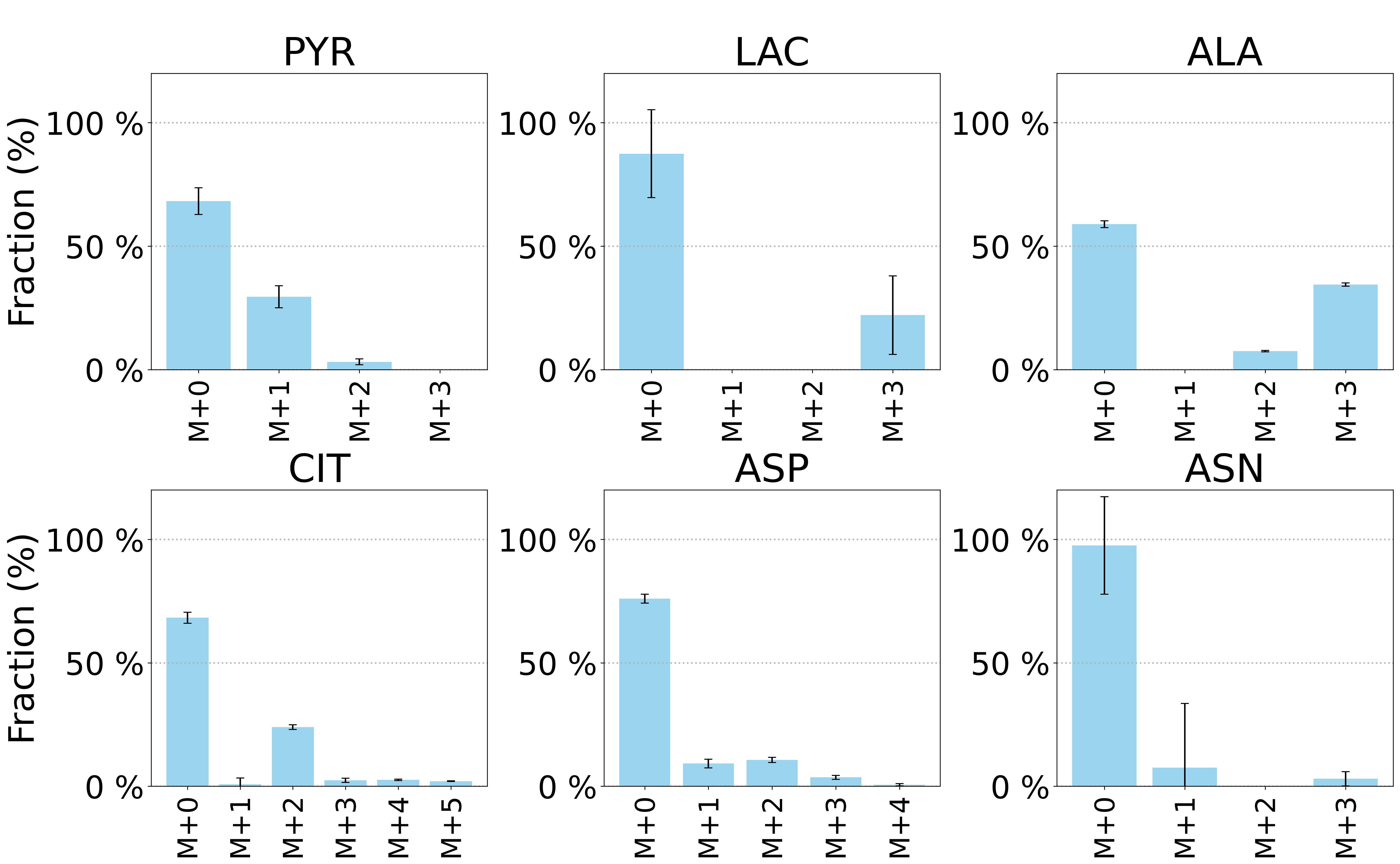}
    \caption{Intracellular MID data for the Static Pyruvate HGLL culture condition at 48~h. MID data were corrected for natural isotopic abundance. Metabolite abbreviations are: PYR (pyruvate), LAC (lactate), ALA (alanine), CIT (citrate), ASP (aspartate), and ASN (asparagine).  {Error bars represent the standard deviation across biological replicates.} A complete list of abbreviations is provided in Supplementary Materials Table S1. The intracellular MIDs for the LGLL, HGLL, and LGHL static pyruvate cultures are provided in the Supplementary Materials (Fig. S1-S3).}
    \label{fig:mid}
\end{figure*}

Across all four Static Pyruvate culture conditions, only low levels of $^{13}$C enrichment were detected in glucose and in the TCA‑cycle intermediates $\alpha$-ketoglutarate, fumarate, succinate, and malate, likely due to detection limitations. For this reason, these metabolites were omitted from the Mass isotopomer distributions (MIDs) shown in Fig.~\ref{fig:mid} for clarity. In contrast, substantial $^{13}$C labeling was observed in lactate and alanine. Previous studies indicate that lactate is derived exclusively from cytosolic pyruvate, whereas alanine can originate from both cytosolic and mitochondrial pools \citep{Varum2011Energy, Buescher2015, Vacanti2014}.  {These observations suggest that most of the exogenously supplied pyruvate remained in the cytosol, with limited incorporation into downstream TCA-cycle intermediates. Additionally, unlabeled glutamine and glutamate supplied in the medium continue to feed the TCA cycle, further diluting $^{13}$C enrichment. Intracellular compartmentalization between cytosolic and mitochondrial pyruvate pools may also contribute to the observed labeling patterns. However, the low enrichment observed in several TCA-cycle intermediates may additionally reflect analytical detection limitations, making it difficult to distinguish between compartmentalization effects and measurement constraints. Therefore, the current isotope-labeling results support limited apparent incorporation of exogenous pyruvate into the TCA cycle but do not permit definitive attribution of the underlying mechanism.} The unlabeled fractions of lactate and alanine likely arise from unlabeled glucose and, to a lesser extent, from the residual unlabeled pyruvate present in the medium.

\subsection{Aggregate Size and Dynamic Distribution}
\label{subsec:Aggregate Size and Dynamic Distribution}
%Profiles}
To support multiscale mechanistic modeling of iPSC aggregate cultures, changes in aggregate size distributions were quantified over the 3‑day culture period. Fig.~\ref{fig:aggregate_combined}A shows representative images of aggregates over the 3‑day culture period, while Fig.~\ref{fig:aggregate_combined}B presents the corresponding violin plots of size distributions and Fig.~\ref{fig:aggregate_combined}C summarizes aggregate counts.
As expected, aggregate size increased substantially over time, from
$265.4 \pm 5.4~\mu\mathrm{m}$ to $412.9 \pm 6.4~\mu\mathrm{m}$ on day 3. The images in Fig.~\ref{fig:aggregate_combined}A illustrate this progression: early aggregates (D1–D2) are smaller and relatively uniform, reflecting initial cluster formation, whereas aggregates at later stages (D3) appear larger and darker, consistent with increased cell density and ongoing proliferation.  The violin plots in Fig.~\ref{fig:aggregate_combined}B corroborate this trend, showing a clear upward shift in size distributions and increasing spread over time, indicating greater heterogeneity in aggregate growth.

In contrast, aggregate counts initially decreased from Day 1 to Day 2 and remained lower than the initial value on Day 3 (Fig.~\ref{fig:aggregate_combined}C). This inverse relationship between aggregate size and aggregate number suggests that growth is driven not only by intracellular proliferation but also by aggregate–aggregate interactions, such as merging or coalescence. As smaller aggregates fuse or expand into larger structures, the culture transitions from many small aggregates to fewer, larger ones. 

This structural evolution has significant functional consequences. With increasing aggregate size, diffusion limitations in oxygen and nutrient transport become more severe, giving rise to substantial spatial and metabolic heterogeneity. Accurately capturing this heterogeneity is essential for multiscale mechanistic foundation modeling to systematically represent iPSC cultures across scales.

\subsection{Redox Measurements for Aggregate Culture}
\begin{figure*}[htbp]
    \centering
    \includegraphics[width=0.8\linewidth]{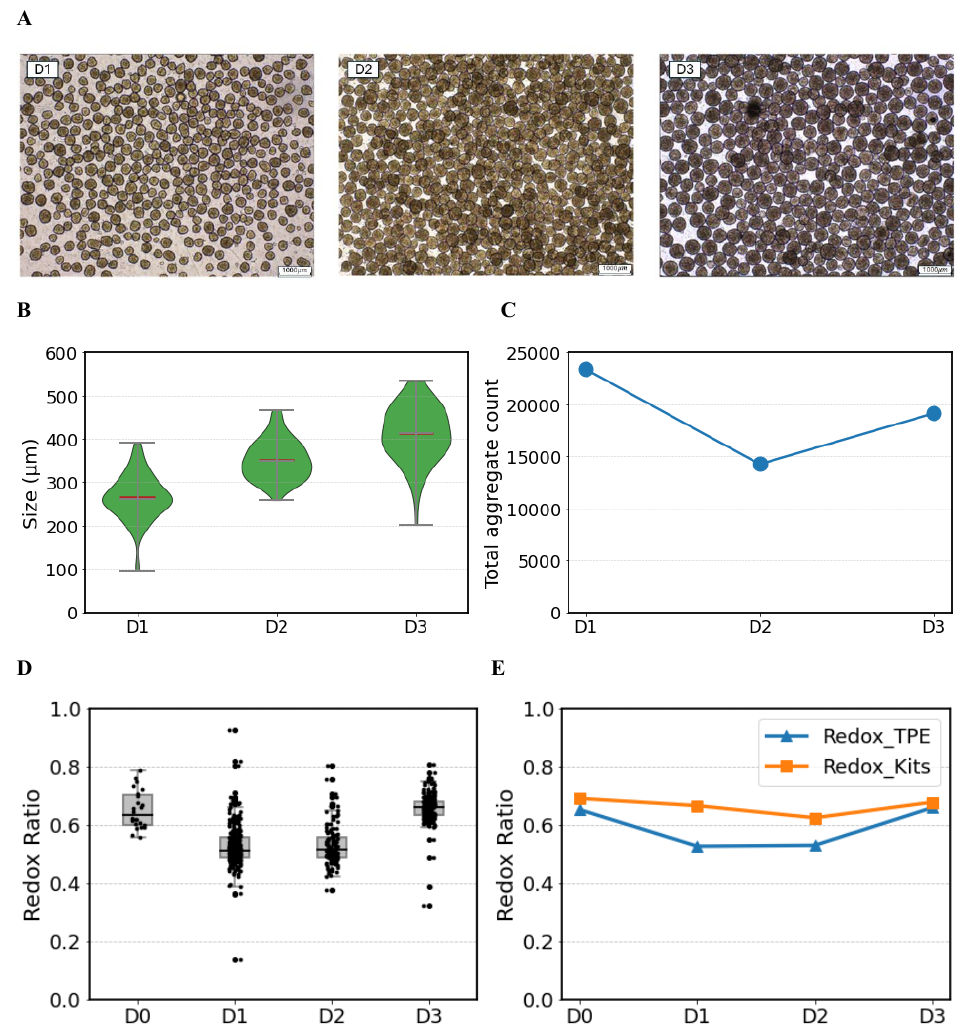}
    \caption{\textbf{Aggregate growth, size distribution, and redox dynamics of iPSC aggregate cultures.}
(\textbf{A}) Representative brightfield images of iPSC aggregates from Day~1 to Day~3 illustrating aggregate growth and morphological evolution over time. The scale bar (bottom right) corresponds to 1000~$\mu$m.
(\textbf{B}) Violin plots showing the distribution of iPSC aggregate sizes across first three days of culture. Green violins represent the kernel density of aggregate size measurements at each time point. Red horizontal bars denote the mean aggregate size, while gray horizontal bars indicate the median. Gray vertical bars and whiskers show the minimum and maximum observed aggregate sizes. Day~3 corresponds to measurements taken prior to feeding, and aggregate size distributions were similar before and after feeding.
(\textbf{C}) Total aggregate counts measured from Day~1 to Day~3.
(\textbf{D}) Distribution of redox ratios measured in aggregate cultures from Day~0 to Day~3 using the TPE sensor. Boxplots show the median and interquartile range, with whiskers indicating the full data range; individual replicate measurements are overlaid as points.
(\textbf{E}) Temporal trends in redox ratios measured by a non-destructive TPE-based method, FAD/(FAD+NAD(P)H), and by an offline enzymatic assay, NAD$^+$/ (NAD$^+$+NADH), demonstrating consistent redox dynamics across measurement modalities. The TPE-based values represent averages calculated over both the measurement period and the analyzed cell population.
}

    \label{fig:aggregate_combined}
\end{figure*}

To quantify redox levels in the aggregate cultures, two complementary methods were used. The enzymatic assay requires cell disruption, after which NAD$^+$ and 
NADH concentrations are measured and used to calculate the redox ratio:
\(\frac{\mathrm{NAD}^+}{\mathrm{NAD}^+ + \mathrm{NADH}}.\) In contrast, the two‑photon excitation (TPE) sensor enables nondestructive measurement of FAD and NAD(P)H autofluorescence signals, from which the optical redox ratio is calculated as:
\(\frac{\mathrm{FAD}}{\mathrm{FAD} + \mathrm{NAD(P)H}}.\) Figs.~\ref{fig:aggregate_combined}D and E show the redox ratios for the aggregate cultures over the 3‑day period. Overall, the redox dynamics captured by the real‑time TPE sensor aligned well with the offline enzymatic measurements, supporting the sensor’s reliability for monitoring metabolic state in iPSC aggregates.

Fig.~\ref{fig:aggregate_combined}E shows the temporal trends of the average redox ratios obtained from 
the TPE sensor and the enzymatic kit measurements. Both methods exhibit similar overall trajectories. The redox ratio decreases from Day 0 to Day~2, followed by a gradual increase from Day 2 to Day~3. This pattern suggests an initial shift toward a reduced metabolic state associated with increased NAD(P)H relative to FAD or NAD$^+$ during early aggregate 
formation and rapid proliferation period, followed by a gradual increase from Days 2–3 as the culture stabilizes and shifts toward a more oxidized state. The consistent trends between the two measurement approaches further validate the TPE sensor as a robust, nondestructive tool for real‑time redox monitoring in iPSC aggregate cultures.

\section{Multi-Scale Model Development}
\label{sec:aggregateModeling}
Based on experimentally characterized iPSC growth kinetics, extracellular metabolite profiles, mass isotopomer distribution (MID) data, intracellular redox measurements, and aggregate size distributions from both static and aggregate cultures, we developed a multiscale mechanistic modeling framework. This framework builds on our previously established single‑cell metabolic model \citep{Wang2024} and extends it to explicitly incorporate intracellular redox dynamics and regulatory interactions associated with pyruvate metabolism. Using this enhanced single‑cell model as the core module, we further integrated population‑balance and reaction–diffusion formulations to represent aggregate formation, transport processes, and intra‑aggregate nutrient gradients. The resulting modular framework was calibrated and validated against the full suite of experimental datasets, enabling quantitative prediction of iPSC metabolic behavior across distinct culture configurations. Details of the single-cell model are presented in Section~\ref{sec:monolayerModeling}, followed by the population balance model (Section~\ref{subsec:aggregation}) and the reaction–diffusion model (Section~\ref{subsec:RDM}).

\begin{figure*}[htbp]
    \centering
    \includegraphics[width=0.96\linewidth]{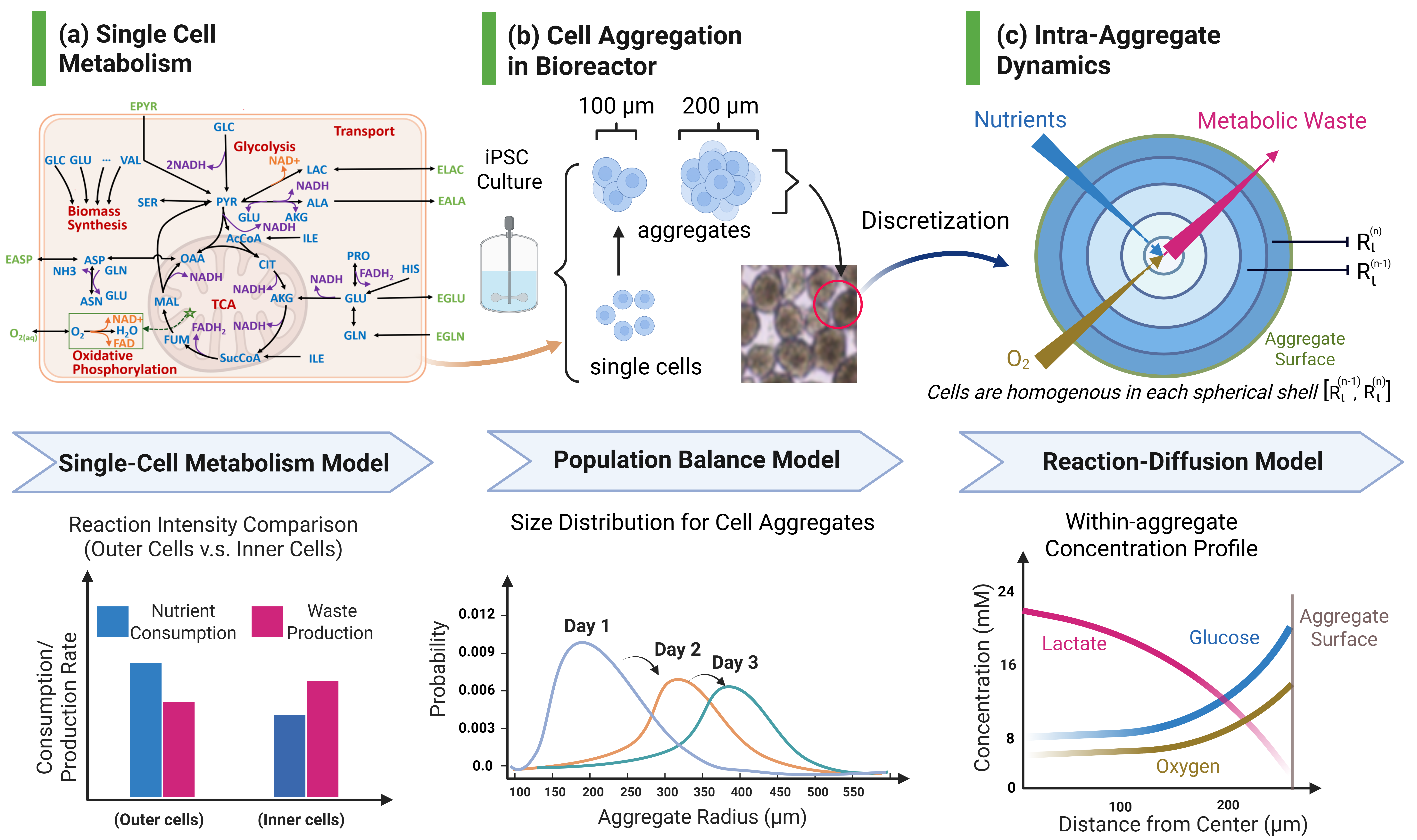}
    \caption{ {Schematic of the proposed multiscale modeling framework integrating single-cell metabolism, population balance modeling (PBM), and reaction–diffusion transport for iPSC aggregate cultures. 
   Intracellular metabolic responses are captured using a single-cell metabolic network, while aggregate formation and size distribution are described by PBM. Spatial gradients of nutrients, metabolites, and oxygen within aggregates are resolved via reaction–diffusion modeling, enabling prediction of heterogeneous microenvironments and the resulting variability in cellular behaviors across aggregate cultures.}
}

    \label{fig:multiscale}
\end{figure*}

\begin{sloppypar}
The resulting multiscale mechanistic framework, referred to as the Biological System-of-Systems (Bio-SoS) model \citep{zheng2024stochastic}, is illustrated in Fig. \ref{fig:multiscale} and comprises three interconnected modules:
\begin{enumerate}
    \item \textbf{Single-Cell Metabolism Model:} Characterizes the coupled metabolic and redox networks that govern cellular responses to spatially heterogeneous micro-environments.
    \item \textbf{Population Balance Model:} Captures cell–cell interactions and the dynamic evolution of iPSC aggregate size distributions in bioreactor cultures.
    \item \textbf{Reaction-Diffusion Model:} Describes intra-aggregate transport dynamics and spatial gradients of nutrients and metabolites within aggregates.
\end{enumerate}

\end{sloppypar}

The developed multiscale mechanistic model, with its modular design, enables the integration of heterogeneous data from both monolayer and aggregate cultures while capturing metabolic and spatial heterogeneity. This framework can guide robust culture optimization, improve production consistency and interoperability, and accelerate iPSC manufacturing scale‑up.

\subsection{Single‑Cell Metabolic Model Extension}
\label{sec:monolayerModeling}
\begin{figure*}[t]
    \centering
    \includegraphics[width=0.73\textwidth]{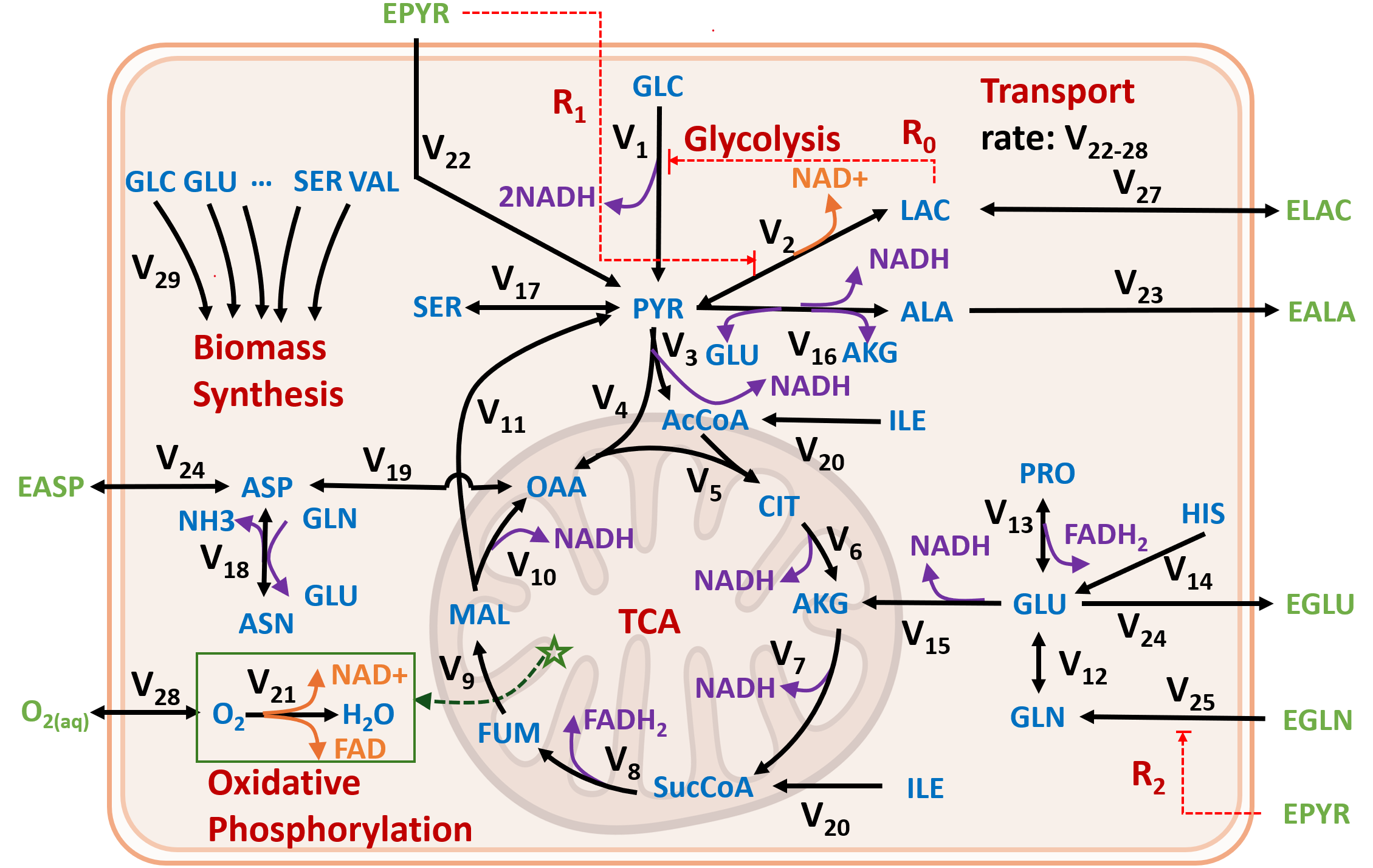}
    \caption{Metabolic network for iPSC including regulatory reactions. Intracellular metabolites are shown in blue, while extracellular metabolites are shown in green and denoted with an ``E'' prefix. Reaction fluxes are represented by black arrows, with transport reactions connecting intracellular and extracellular species. Energy-generating steps are highlighted in purple, whereas energy-consuming steps are indicated in orange. Oxidative phosphorylation in the mitochondria, associated with NADH oxidation and oxygen consumption, is indicated by the green star. Reactions involving NAD$^{+}$/NADH and FAD/FADH$_2$ are explicitly annotated to capture cellular redox balance. The network integrates glycolysis, the TCA cycle, oxidative phosphorylation, amino acid metabolism, and biomass synthesis. EPYR in Static Pyruvate culture represents both labeled and unlabeled extracellular pyruvate.
 }
    \label{fig:metabolic_network}
\end{figure*}

\begin{sloppypar}
Building on the results reported in \cite{Wang2024}, this section extends the mechanistic single‑cell metabolic model developed for 2D monolayer iPSC cultures. This enhanced model serves as the core computational module within the modular multiscale framework for 3D aggregate cultures.
Because 2D monolayer cultures experience relatively homogeneous environmental conditions, the associated single‑cell metabolic model—schematically illustrated in Fig.~\ref{fig:metabolic_network}—builds on previously established frameworks \citep{ghorbaniaghdam2014analyzing, Wang2024}.  {Metabolites exhibiting significant temporal changes over the 48‑h exponential growth phase \cite{Odenwelder2021} were identified via linear regression slope tests ($p<0.05$) and incorporated into model development. Proline (PRO) was retained despite not meeting this criterion, as its distinct behavior in aggregate cultures suggests a unique role in aggregate‑specific metabolism.}

Relative to the prior iPSC metabolic regulatory network model \citep{Wang2024}, the present formulation explicitly incorporates intracellular redox reactions through the inclusion of redox cofactors (NAD$^{+}$/NADH and FAD/FADH$_2$) and a simplified representation of oxidative phosphorylation. These extensions enable the model to capture the experimentally observed coupling between central carbon metabolism and cellular redox balance.

To represent the inhibitory effect of lactate on glucose uptake, a lactate‑dependent regulatory element was incorporated following Odenwelder et al. \citep{Odenwelder2021} and is denoted as R0 in Fig.~\ref{fig:metabolic_network}. In addition, motivated by experimental observations from the present study, two additional regulatory mechanisms (R1 and R2) were introduced to account for the effects of pyruvate on cellular metabolism.
\end{sloppypar}

 \vspace{0.1in}

\noindent \textbf{R1: Pyruvate inhibition on lactate production} \citep{Rao2021} (R1 in Fig. \ref{fig:metabolic_network}). Recent experimental evidence demonstrates that elevated extracellular pyruvate can inhibit lactate dehydrogenase (LDH) activity in living cells through an MCT1-dependent mechanism \citep{Rao2021}. Pyruvate and lactate share the same monocarboxylate transporters (MCTs), and high extracellular pyruvate concentrations increase intracellular pyruvate levels through MCT1-mediated transport. The resulting rise in intracellular pyruvate shifts the LDH equilibrium, suppressing the forward conversion of pyruvate to lactate by perturbing the NADH/NAD$^{+}$ redox balance and competing with lactate export. Consequently, excess pyruvate reduces net lactate production and redirects carbon flux toward mitochondrial oxidation pathways.

To capture this regulatory effect, an inhibitory term dependent on extracellular pyruvate concentration (EPYR) is incorporated into the LDH rate expression: 
\begin{align}
v(\mathrm{LDH})
&= v_{\mathrm{maxfLDH}}
\frac{\mathrm{PYR}}{K_{m\mathrm{PYR}}+\mathrm{PYR}}
\nonumber \times
\frac{\mathrm{NADH}/\mathrm{NAD}}
     {K_{m\mathrm{NADHtoNAD}}+\mathrm{NADH}/\mathrm{NAD}}
\nonumber\times
\frac{K_{i\mathrm{EPYR}}}{K_{i\mathrm{EPYR}}+\mathrm{EPYR}}
\nonumber\\
&\quad -
v_{\mathrm{maxrLDH}}
\frac{\mathrm{LAC}}{K_{m\mathrm{LAC}}+\mathrm{LAC}}
\nonumber\times
\frac{\mathrm{NAD}/\mathrm{NADH}}
     {K_{m\mathrm{NADtoNADH}}+\mathrm{NAD}/\mathrm{NADH}} .\nonumber
\label{eq:R1}
\end{align}
This term reduces the effective LDH activity as extracellular pyruvate accumulates, capturing the experimentally observed suppression of lactate production under high‑pyruvate conditions.

\vspace{0.1in}
\noindent \textbf{R2: Pyruvate inhibition on glutamine consumption} \citep{Yang2014} (R2 in Fig. \ref{fig:metabolic_network}). 
Glutamine is a major anaplerotic substrate that replenishes tricarboxylic acid (TCA) cycle intermediates and supports biosynthesis in proliferating cells. Previous studies have shown that glutamine oxidation becomes particularly important when mitochondrial pyruvate transport is limited, enabling cells to sustain TCA cycle activity in the absence of sufficient pyruvate‑derived carbon \citep{Yang2014}. Under these conditions, glutamine-derived carbon is converted to $\alpha$-ketoglutarate and fuels downstream TCA cycle reactions. Conversely, when pyruvate availability is high, mitochondrial
pyruvate oxidation can directly provide acetyl-CoA to the TCA cycle, reducing the reliance on glutamine as an anaplerotic substrate. 
To capture this metabolic regulation, an inhibitory term dependent on extracellular pyruvate concentration (EPYR) is incorporated into the glutamine consumption rate expression:
\begin{align}\nonumber
v(\mathrm{GLNS})
&= v_{\mathrm{maxfGLNS}}
\frac{\mathrm{GLN}}{K_{m\mathrm{GLN}}+\mathrm{GLN}}
\times
\frac{K_{i\mathrm{EPYR}}}{K_{i\mathrm{EPYR}}+\mathrm{EPYR}}
-
v_{\mathrm{maxrGLNS}}
\frac{\mathrm{GLU}}{K_{m\mathrm{GLU}}+\mathrm{GLU}}
\times
\frac{\mathrm{NH_4}}{K_{m\mathrm{NH4}}+\mathrm{NH_4}} .
% \label{eq:R2}
\end{align}

Overall, the single-cell metabolism model contains 29 reactions, each associated with a corresponding flux rate. All reactions are listed in the Supplementary Materials (Table~S2), along with the corresponding flux‑rate expressions in Table~S3. In addition to incorporating the expanded redox network, the model extends the previous iPSC mechanistic framework \citep{Wang2024} by introducing a biomass‑associated growth rate (see Equation~(29) in Table~S3). The stoichiometric coefficients for amino acid consumption in the biomass equation were derived from the amino acid composition reported in \cite{xu2023high}, enabling a more accurate representation of iPSC growth kinetics and amino acid consumption.

{
To capture intrinsic bioprocess variability at the molecular level, we develop a stochastic single-cell metabolic model based on a stochastic reaction-network formulation \citep{anderson2011continuous,zheng2024stochastic}. Fluxes $\mathbf{v}$ represent the regulatory mechanisms listed in Table~S3, with reaction occurrences over a small interval $\Delta t$ modeled as Poisson processes with expected counts $\mathbf{v}\Delta t$. Resulting reaction events update intracellular metabolite concentrations via the stoichiometric matrix $\mathbf{N}$ (Table~S2). This formulation preserves the mechanistic structure defined by $\mathbf{N}$ and $\mathbf{v}$ while capturing stochastic fluctuations in intracellular metabolism.
}

\subsection{Cell–Cell Interactions and Aggregation Dynamics}
\label{subsec:aggregation}

To characterize the formation and evolution of iPSC aggregates, we employed a population balance model (PBM) \citep{ aldous1999deterministic, ramkrishna2014population, wu2014oxygen} to describe cell–cell interactions and aggregation dynamics during aggregate culture. In these systems, individual cells and small clusters progressively collide, adhere, and proliferate, giving rise to multicellular aggregates whose size distribution changes over time (see Section~\ref{subsec:Aggregate Size and Dynamic Distribution}). These dynamics arise from the interplay of biological processes—such as cell adhesion, proliferation, and extracellular matrix deposition—and physical mechanisms including aggregate–aggregate coalescence. The PBM offers a quantitative framework for capturing these effects by tracking the time-dependent evolution of the aggregate size distribution.

Let $\phi(x,t)$ denote the number density of aggregates of size $x$ at time $t$. The temporal evolution of the size distribution is governed by the population balance equation:
\begin{equation}
\label{eq.PBM}
\begin{aligned}\nonumber
\frac{\partial \phi(x,t)}{\partial t}
&= \frac{1}{2} \int_{x_0}^{x} \phi(x_c,t)\phi(x',t)K(x_c|x') \, dx' - \int_{x_0}^{\infty} \phi(x,t)\phi(x',t)K(x|x') \, dx' - \frac{\partial}{\partial x}
\left[
\phi(x,t)\frac{\partial x}{\partial t}
\right].
\end{aligned}
\end{equation}
The three terms on the right-hand side represent the fundamental mechanisms shaping aggregate size evolution.
The first term captures the \textit{formation} of aggregates of size $x$  through the merger of two smaller clusters of sizes $x'$ and $x_c = x - x'$. The product $\phi(x_c,t)\phi(x',t)$ reflects their interaction frequency.
The second term describes the \textit{loss} of aggregates of size $x$ as they combine with other clusters to form larger aggregates.
The third term accounts for  \textit{size growth} driven by cell proliferation within aggregates, modeled as
$\frac{\partial x}{\partial t}
=
\alpha_G \, x
\log\left(\frac{M}{x}\right)$,
where $\alpha_G$ is proportional to the cellular growth rate and $M$ represents the maximum attainable aggregate size.

The aggregation kernel governing interactions between aggregates is defined as 
\begin{equation}\nonumber
K(x|x') =
k \cdot
\exp\left(
-k_1\left(\frac{x+x'}{2}\right)^{a}
\right)
\left(x^{\frac{1}{3}} + x'^{\frac{1}{3}}\right)^{7/3}
\end{equation} following \cite{aldous1999deterministic}.
In this formulation, the parameter $k$ represents the baseline hydrodynamic collision frequency and serves as the overall aggregation rate constant. The exponential term captures the size‑dependent decline in adhesion probability due to surface‑related constraints;
%due to steric hindrance, reduced surface accessibility, or biological surface effects; 
here $k_1$ controls the magnitude of this size‑dependent inhibition, and $a$ determines the sensitivity of adhesion efficiency to aggregate size. The geometric factor $\left(x^{\frac{1}{3}} + x'^{\frac{1}{3}}\right)^{7/3}$ 
encodes how the effective collision cross‑section scales with the sizes of the interacting aggregates, reflecting classical geometric considerations in cluster–cluster aggregation.

\subsection{Coupled Reaction and Diffusion Dynamics}
\label{subsec:RDM}
%\paragraph{\textbf{Reaction-Diffusion Model (RDM).}}

To characterize spatial gradients of nutrients and metabolites within aggregates, a reaction--diffusion model \citep{tosaka1982analysis, wu2014oxygen} was employed to describe the transport and consumption of extracellular metabolites. Aggregates are approximated as spherical cell clusters with radial symmetry. The concentration of metabolite $i$ within the aggregate satisfies
\begin{equation}\nonumber
\frac{\partial c_i}{\partial t}
=
\frac{D_i}{r^2}
\frac{\partial}{\partial r}
\left(
r^2
\frac{\partial c_i}{\partial r}
\right)
+
\rho_i(c,s)~,
\end{equation}
where $c_i(r,t)$ denotes the concentration of metabolite $i$ at radial position $r \in [0,R]$ and time $t$, with $R$ denoting the aggregate radius. Here, $D_i$ is the effective diffusion coefficient of metabolite $i$ within the aggregate, and $\rho_i(c,s)$ represents the local metabolic reaction rate determined by the intracellular metabolic model (see Section~\ref{sec:monolayerModeling}), where $c$ denotes the vector of extracellular metabolite concentrations and $s$ represents the intracellular metabolic states governing cellular uptake and secretion.

 {Through coupling with the single‑cell metabolic model, the reaction–diffusion model provides spatial gradients of nutrients, metabolites, and oxygen, giving rise to heterogeneous extracellular microenvironments within aggregates. These local conditions directly regulate intracellular reaction rates, leading to spatially varying glycolytic activity, TCA‑cycle fluxes, amino acid metabolism, and redox states across the aggregate radius. Cellular consumption of nutrients and secretion of metabolic byproducts, in turn, reshape the local extracellular environment, establishing a dynamic feedback loop between intracellular metabolism and extracellular transport. As aggregates grow and diffusion limitations intensify, this feedback amplifies metabolic heterogeneity among cells within the aggregate.
}

\section{Model Fitting and Validation}
\label{sec:modelFitting}

\begin{figure*}[htbp]
    \centering
    \includegraphics[width=0.8\linewidth]{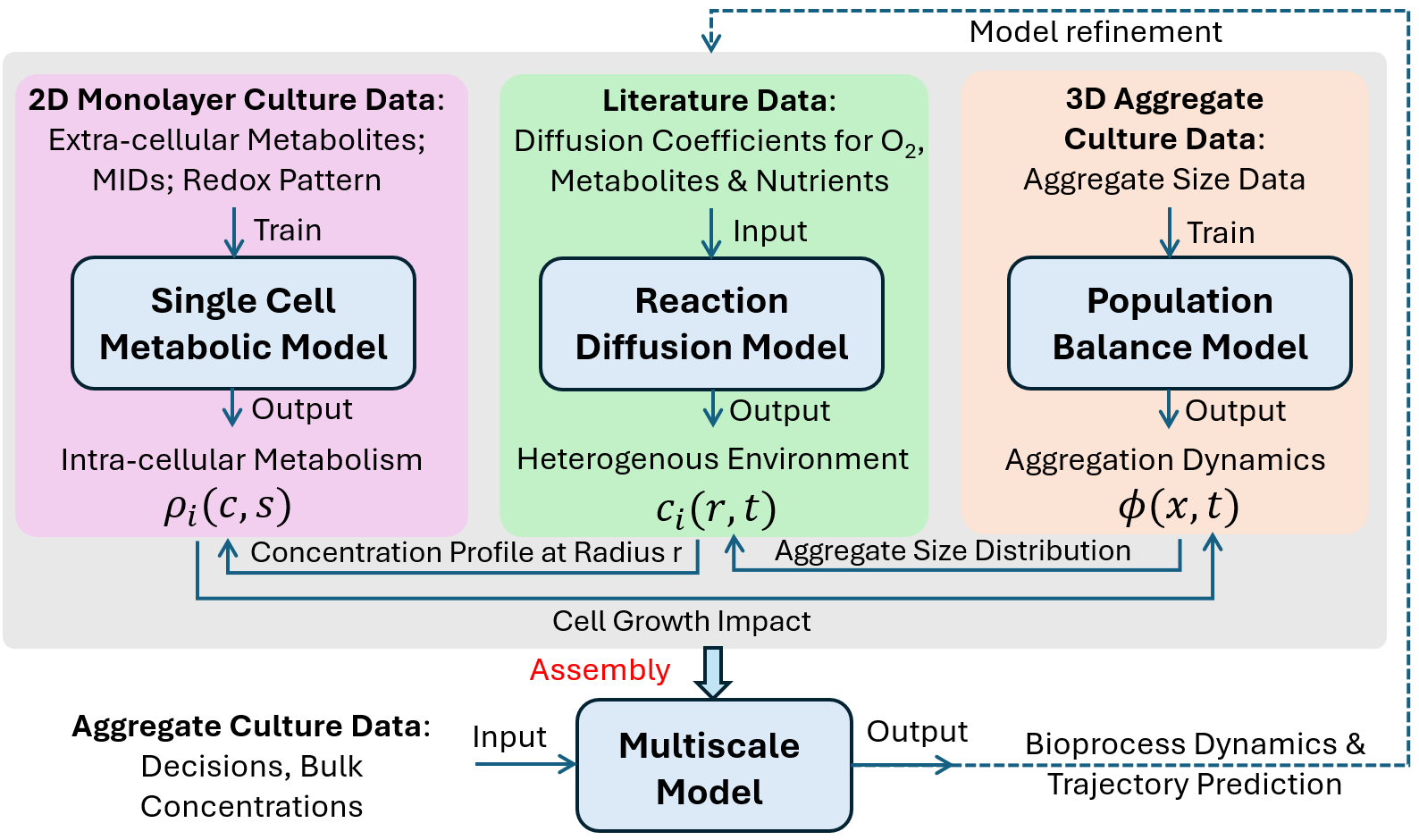}
    \caption{ {Data flow of the proposed multiscale modeling framework integrating single-cell metabolism, reaction–diffusion transport, and population balance modeling. Experimental measurements and literature-derived parameters provide inputs to each module, which are coupled through shared bioprocess state variables to construct an integrated model for predicting iPSC aggregate culture dynamics.
    }
}

    \label{fig:multiscale_structure}
\end{figure*}

The multiscale foundation model integrates three components: (i) a single‑cell mechanistic model, (ii)  a population balance model describing the aggregation process, and (iii) a diffusion–reaction module that captures intra‑aggregate transport and reaction dynamics of nutrients and metabolites.
We iteratively leveraged heterogeneous measurements from both 2D monolayer and 3D aggregate cultures to calibrate the modules of the multiscale mechanistic foundation model, as detailed in Fig. \ref{fig:multiscale_structure}.

\vspace{0.1in}
\noindent \textbf{(1) Single-cell metabolism model fitting.}
All eight monolayer datasets (four Historic Static conditions \cite{Odenwelder2021} and four Static Pyruvate conditions) were used to fit the cell metabolic kinetic model parameters described in Table~S4. Parameter estimation was performed by minimizing the mean squared error (MSE) between model predictions and experimental measurements: %i.e.,
\begin{equation}\label{eq:MSE}\nonumber
\mathrm{MSE} = \frac{1}{K I} 
\sum_{k=1}^{K} \sum_{i=1}^{I} \frac{1}{T} \sum_{t \in \mathcal{T}} 
\left( y_{i,t}^{(k)} - \hat{y}_{i,t}^{(k)} \right)^{2}, 
\end{equation}
where $y_{i,t}^{(k)}$ is the experimental measurement of state variable $i$ at time $t$ for dataset $k$, and $\hat{y}_{i,t}^{(k)}$ is the corresponding model prediction. The measurement time sets $\mathcal{T} = \{12,24,36,48\}$ or $\mathcal{T} = \{24,48\}$ depend on the specific experimental study $k$. Here, $K$ denotes the total number of datasets, $I$ is the total number of state variables measured, and $T$ the total number of measurements. The estimated single‑cell mechanistic parameter values are provided in the Supplementary Materials (Table~S4).

\vspace{0.1in}
\noindent \textbf{(2) Population balance model and training.} Aggregate size measurements were used to calibrate a population balance model (PBM) that characterizes aggregation dynamics. The predicted aggregate distribution $\hat{\phi}(x,t)$ and the empirical observations on iPSC aggregate size distribution $\phi(x,t)$ from day~1 to day~3 were compared by minimizing the KL divergence:
\begin{equation}\nonumber
\min D_{KL}(\phi \parallel \hat{\phi}) 
= \min \sum_{t \in \mathcal{T}} \phi(x,t) 
\log \left( \frac{\phi(x,t)}{\hat{\phi}(x,t)} \right) dx,
\end{equation}
where $\mathcal{T} = \{24,48,72\}$ representing days~1 to 3.

\vspace{0.1in}
\noindent \textbf{(3) Reaction-diffusion module and training.}
The reaction–diffusion module describes the transport and consumption of nutrients and metabolic wastes, enabling quantification of the spatially heterogeneous microenvironment arising from cell–cell interactions within aggregates. This module models the spatiotemporal evolution of key extracellular components, including nutrients and metabolites such as glucose and lactate. Aggregates are approximated as spherical cell clusters with radial symmetry, assuming isotropic diffusion along the radial direction. Detailed model implementation are provided in \citep{zheng2024stochastic}.

For training the reaction–diffusion module, the diffusion coefficients of each substrate (initial values provided in Supplementary Table~S5) were perturbed to minimize the mean squared error (MSE) between model predictions and extracellular metabolite measurements from the aggregate culture dataset. The loss function follows MSE defined in Section~\ref{sec:modelFitting}, with  {$K=2$ corresponding to two biological replicates} and the measurement time set $\mathcal{T} = {24, 48, 72}$.

\section{Model Prediction Performance}
Following model development and calibration, the predictive capability of the proposed framework was evaluated across multiple biological scales and culture configurations. Model performance was assessed by comparing predictions against experimental measurements of extracellular metabolite dynamics, intracellular MIDs, redox responses, aggregate size evolution, and aggregate‑scale metabolic behavior. These analyses evaluated both the quantitative agreement between model predictions and experimental observations and the framework’s ability to generalize across manufacturing scales exhibiting different degrees of heterogeneity.

Subsections~\ref{subsec:monlayerPrediction} and \ref{subsec:aggregatePrediction} present prediction results for 2D monolayer and 3D aggregate cultures, respectively. For monolayer systems, the evaluation focuses on extracellular metabolite dynamics, isotopic labeling behavior, and intracellular redox responses. For aggregate cultures, the assessment further examines the ability of the multiscale framework to reproduce aggregate growth, size distribution evolution, and extracellular metabolic profiles arising from diffusion‑limited microenvironments and cell–cell interactions.

\subsection{Monolayer Model Prediction Performance}
\label{subsec:monlayerPrediction}

This section evaluates the predictive performance of the single‑cell metabolic model under monolayer culture conditions. Because monolayer systems provide a relatively homogeneous extracellular environment, they offer a controlled setting for assessing the model’s ability to capture intracellular metabolic regulation. Model performance is examined from three complementary perspectives: (i) prediction of extracellular metabolite dynamics under diverse nutrient conditions (Section~\ref{subsubsec:Extracellular Metabolite Prediction}), (ii) reproduction of intracellular mass isotopomer distributions from tracer experiments (Section~\ref{subsubsec:MID Validation and Prediction}), and (iii) prediction of intracellular redox responses to environmental perturbations (Section~\ref{subsubsec:Model Prediction on Redox Dynamics}). Together, these analyses provide a systematic validation of the single‑cell metabolic model presented in Section~\ref{sec:monolayerModeling} before extending the evaluation to the spatially heterogeneous aggregate cultures (Sections~\ref{sec:monolayerModeling}-\ref{subsec:RDM}).

\subsubsection{Extracellular Metabolite Prediction}
\label{subsubsec:Extracellular Metabolite Prediction}
\begin{sloppypar}
To evaluate the model’s robustness and predictive performance across different extracellular environments, a leave‑one‑out cross‑validation strategy was employed. In each iteration, the model was trained on seven of the eight culture conditions—four Historic Static conditions \citep{Odenwelder2021} and four Static Pyruvate conditions—and tested on the remaining condition. This procedure enabled us to test the model’s ability to generalize to previously unseen environmental settings, including new combinations of glucose, lactate, and pyruvate concentrations.
\end{sloppypar}

The cross-validation results across eight culture conditions demonstrated that the model achieved reasonably good prediction performance under diverse nutrient and metabolite environments. For the Historic Static cultures, the model consistently captured key metabolic behaviors such as glucose and glutamine consumption, lactate and ammonium accumulation, and amino acid dynamics, as shown in the Supplementary Materials (Figs. S4-S7). Particularly strong agreement was observed in HGHL, HGLL and LGLL conditions, while the LGHL condition showed minor deviations—primarily in the alanine profile (EALA). Nonetheless, overall temporal trends were well reproduced, indicating that the model effectively generalizes to unseen conditions and accurately captures metabolic shifts driven by nutrient depletion or byproduct accumulation.

For the Static Pyruvate cultures, the model showed slightly reduced predictive performance compared to the Historic Static conditions, particularly in capturing alanine dynamics. This discrepancy may arise from unmodeled compartmentalization between cytosolic and mitochondrial pyruvate pools, which influence alanine biosynthesis through distinct pathways. Model predictions for the HGLL Static Pyruvate condition are shown in Fig.~\ref{fig:metaboliteprediction_PYR_HGLL}, with results for the remaining three conditions provided in the Supplementary Materials (Figs.~S8–S10).  {The relatively large prediction interval observed for lactate likely reflects its strong dependence on intracellular redox balance and pyruvate-associated regulation. Because lactate production is governed by the LDH reaction, which directly couples pyruvate metabolism with the NADH/NAD$^{+}$ redox state, small variations in metabolic flux distributions can propagate into comparatively large variations in lactate production.} In addition, the model substantially underestimated cell density in the LGLL condition, which consequently reduced the accuracy of metabolite predictions—even when qualitative trends appeared visually consistent. These limitations may also be partially attributed to the smaller number of biological replicates in the Static Pyruvate datasets (2 replicates) relative to the Historic Static cultures (6 replicates), increasing estimation uncertainty.

 {The predictive performance for individual biological replicates was quantified using the normalized mean absolute error (nMAE), with results summarized in Table~\ref{tab:nmae_validation}. The nMAE was computed as
\begin{equation}\nonumber
\mathrm{nMAE}_i
=
\frac{1}{T}
\sum_{t=1}^{T}
\frac{\left|y_{i,t}-\hat{y}_{i,t}\right|}{\max(y_{i,1},y_{i,2},\ldots,y_{i,T})}
\times 100\%,
\end{equation}
where $y_{i,t}$ and $\hat{y}_{i,t}$ denote the experimental measurement and model prediction of state variable $i$ at time $t$, respectively, and $T$ is the total number of measurements. nMAE was adopted to enable comparison across state variables with differing units and scales, for which normalized metrics are more appropriate than absolute errors \cite{piotrowski2022evaluation}.}

 {
Although several Static Pyruvate replicates exhibited higher nMAE values than Historic Static cultures, the model captured the dominant metabolic trends across all conditions. Parameter uncertainty was further quantified via bootstrap analysis across biological replicates, with median estimates and corresponding 95\% confidence intervals reported in Table~S4. These intervals indicate that parameters are reasonably well constrained despite the limited number of replicates and potential correlations. Overall, these results demonstrate the robustness of the calibrated model and its ability to predict metabolic responses under pyruvate supplementation, supporting its applicability to both standard and modified iPSC culture conditions.} 

\begin{figure}[h]
    \centering    \includegraphics[width=0.85\linewidth]{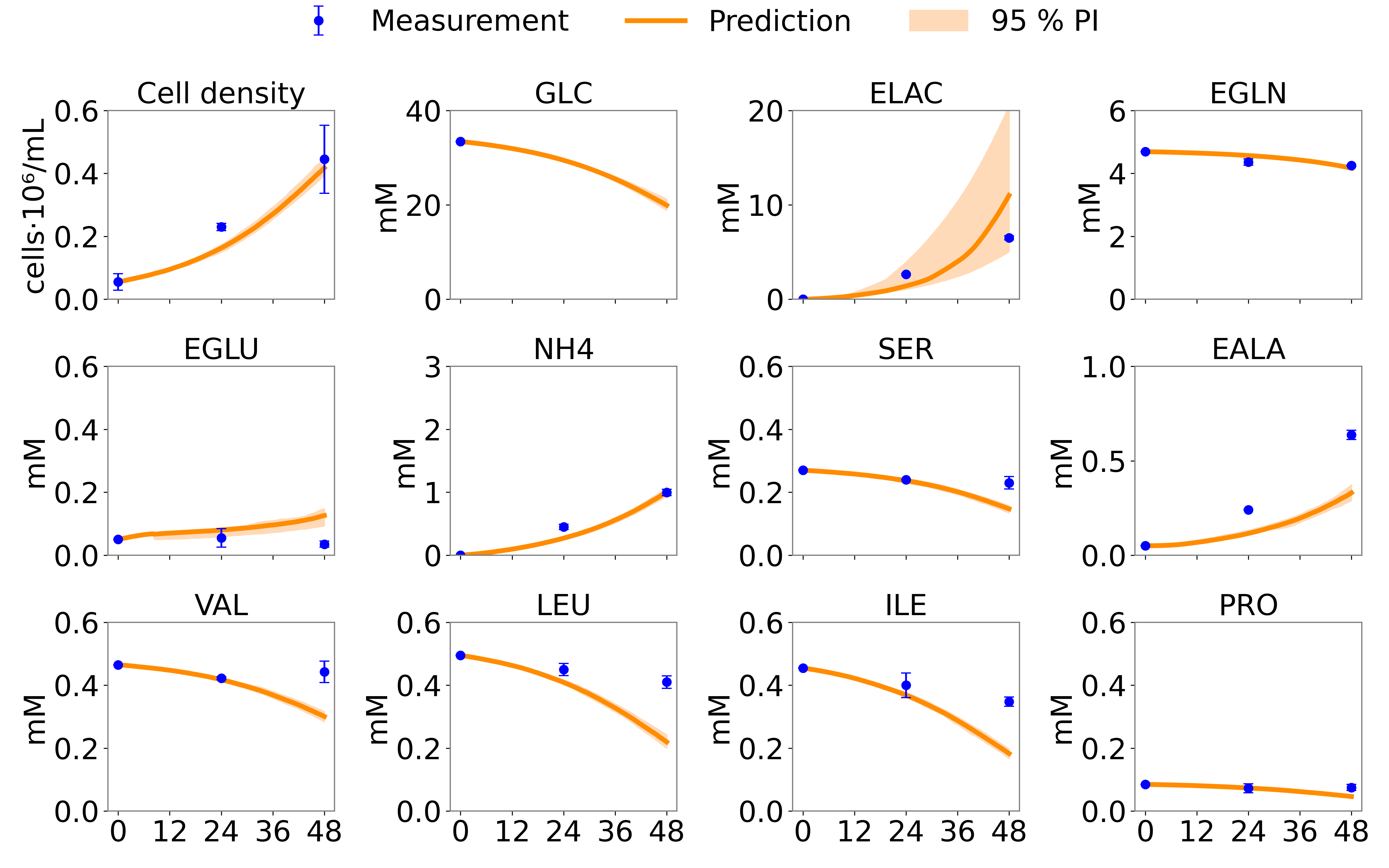}
    \caption{Dynamic model cross-validation for the HGLL static pyruvate condition. The model was trained using all Historic Static culture datasets and the remaining Static Pyruvate culture conditions and then used to predict the HGLL case. Blue dots indicate mean experimental measurements and  error bars represent the standard deviation observed across HGLL static pyruvate culture replicates.  {Orange curves denote the mean model prediction, while the shaded orange regions indicate the 95\% prediction interval.} Cross-validated predictions for all static conditions are provided in the Supplementary Materials (Fig S2–S8).}
    \label{fig:metaboliteprediction_PYR_HGLL}
\end{figure}

\begin{table*}[htbp]
\centering
\caption{Mean nMAE (\%) across metabolites shown in Fig.~\ref{fig:metaboliteprediction_PYR_HGLL} for Historical Static and Static Pyruvate cultures.}
\label{tab:nmae_validation}

\begin{tabular}{lcccccc|cc}
\hline
& \multicolumn{6}{c|}{\textbf{Historical Static}} 
& \multicolumn{2}{c}{\textbf{Static Pyruvate}} \\
\cline{2-9}
\textbf{Condition}
& Rep1 & Rep2 & Rep3 & Rep4 & Rep5 & Rep6
& Rep1 & Rep2 \\
\hline

HGHL
& 10.30 & 11.02 & 11.03 & 13.13 & 11.85 & 11.82
& 8.70 & 14.41 \\
HGLL
& 11.32 & 11.10 & 10.50 & 10.98 & 11.43 & 11.72
& 20.09 & 16.45 \\
LGLL
& 10.67 & 11.16 & 11.09 & 12.68 & 12.34 & 13.13
& 10.56 & 10.84 \\
LGHL
& 12.76 & 12.33 & 14.65 & 13.03 & 13.44 & 13.68
& 18.70 & 16.53 \\
\hline

\end{tabular}
\end{table*}

\subsubsection{MID Validation and Prediction}
\label{subsubsec:MID Validation and Prediction}

\begin{figure*}[t]
    \centering
    \includegraphics[width=0.9\linewidth]{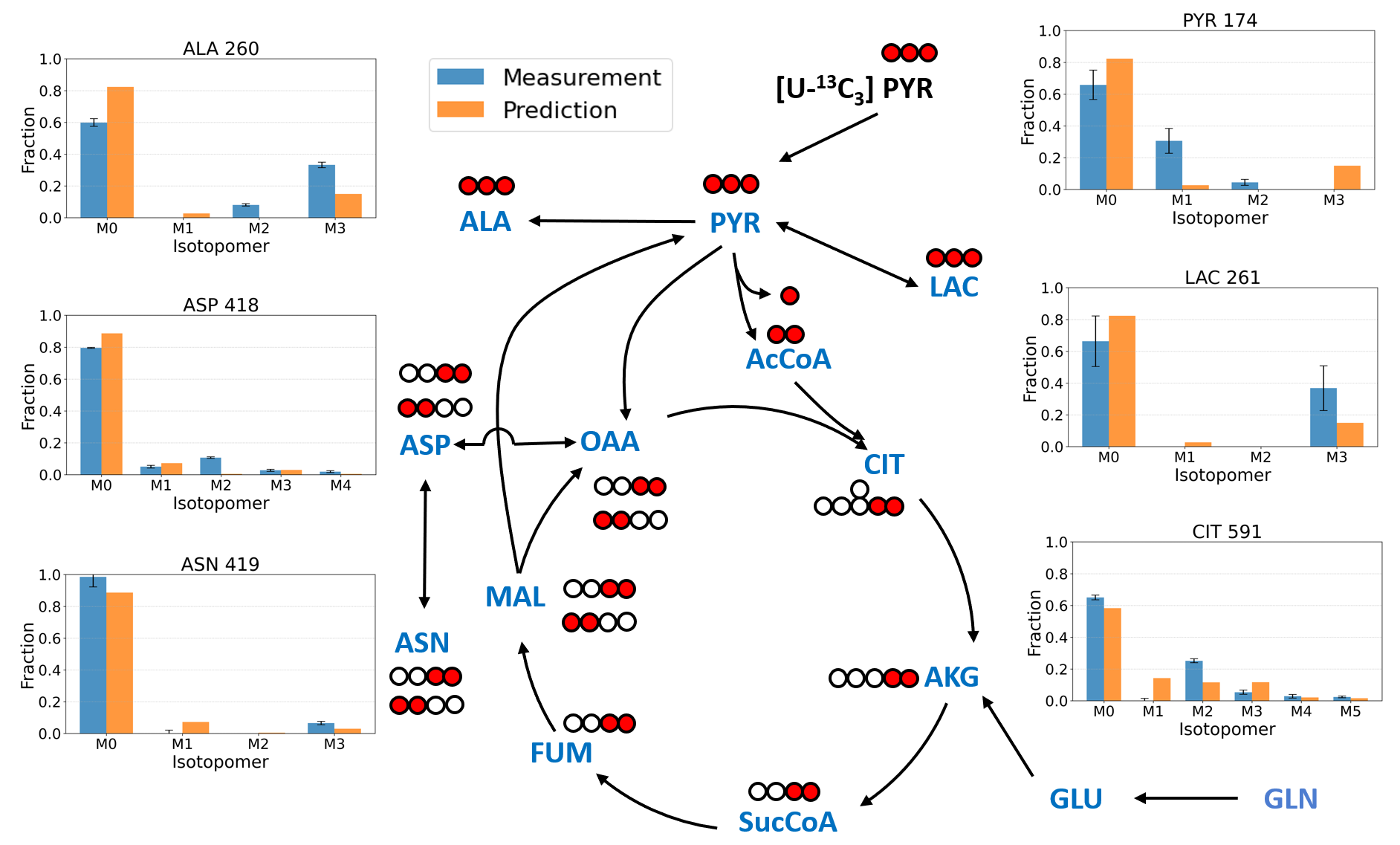}
    \caption{Model prediction of mass isotopomer distributions for glycolytic and TCA cycle metabolites in the Static Pyruvate culture (HGLL) with [U-\(^{13}\)C\(_3\)]pyruvate as the tracer. The central schematic illustrates representative carbon transfer pathways from pyruvate into downstream metabolites through glycolysis, pyruvate metabolism, and the tricarboxylic acid (TCA) cycle. Red circles denote $^{13}$C-labeled carbons originating from the tracer. Measured intracellular MIDs (blue) and predicted (orange) are shown. 
    }
    \label{fig:mid_PYR}
\end{figure*}

\begin{sloppypar}
To evaluate the mechanistic model’s ability to represent central carbon metabolism in iPSCs, simulated MIDs were compared with representative experimental labeling data obtained under the Historic Static HGLL condition using [1,2-\(^{13}\)C\(_2\)] glucose and [U-\(^{13}\)C\(_5\)] glutamine as tracers. After validating model performance against the Historic Static datasets, the model’s predictive capability was further assessed under the Static Pyruvate condition—without any parameter refitting—using [U-\(^{13}\)C\(_3\)] pyruvate as the isotopic tracer. The results are shown in Fig.~\ref{fig:mid_PYR}. The model successfully captures the pyruvate‑driven labeling patterns, including strong M\(+3\) enrichment in pyruvate‑derived metabolite pools and a corresponding reduction in contributions from glucose‑derived labeling. 
\end{sloppypar}

 {Predictive accuracy decreases for isotopomer distributions of downstream TCA-cycle intermediates. This reduced performance may reflect intracellular compartmentalization effects, limited apparent incorporation of exogenous pyruvate into the TCA cycle, and the inherently low signal-to-noise ratios associated with TCA-cycle labeling measurements. Because the observed enrichments in several TCA-cycle intermediates were low, it remains difficult to distinguish between biological compartmentalization effects and measurement limitations. Nevertheless, these results highlight opportunities for improving quantitative prediction of TCA-cycle labeling patterns through more detailed characterization of intracellular pyruvate metabolism and mitochondrial processes.}

\subsubsection{Model Prediction on Redox Dynamics}
\label{subsubsec:Model Prediction on Redox Dynamics}

\begin{figure*}[t]
    \centering
    \includegraphics[width=\textwidth]{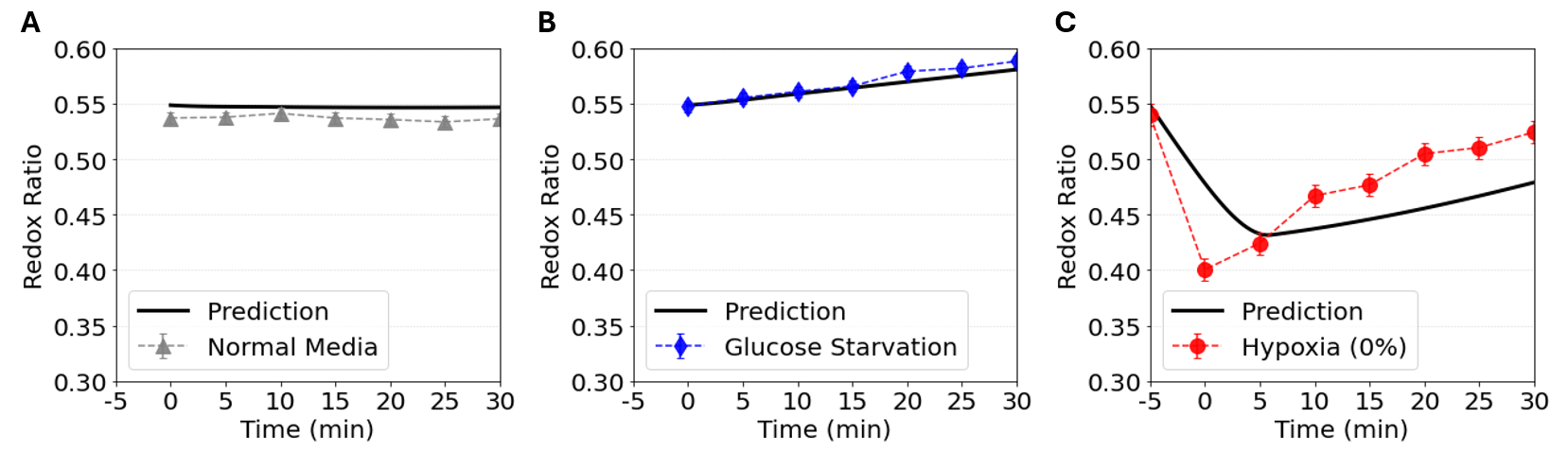}
    \caption{Predicted redox ratio dynamics under different environmental perturbations.
(A) Normal medium, (B) glucose starvation, and (C) hypoxia (0\% O$_2$).
Solid black lines denote model predictions for iPSC, while symbols represent experimental measurements.
Data were reported in \cite{Liu2018}.}

    \label{fig:redox_prediction}
\end{figure*}

To further validate the single-cell metabolic model, we performed simulations to predict intracellular redox dynamics under different culture perturbations.  
Fig.~\ref{fig:redox_prediction} shows the simulated dynamics of the intracellular redox ratio, defined as NAD$^+$/(NAD$^+$ + NADH), across distinct culture conditions. 
Under normal medium conditions, the predicted redox ratio remains relatively stable throughout the simulation, reflecting sustained redox homeostasis (Fig.~\ref{fig:redox_prediction}A). In this regime, NADH generated through glycolysis and the tricarboxylic acid (TCA) cycle is efficiently reoxidized to NAD$^+$ via lactate dehydrogenase (LDH) activity and oxidative phosphorylation (OXPHOS), resulting in a balanced production and consumption of reducing equivalents.

Under glucose starvation, the model predicts a modest increase in the redox ratio (Fig.~\ref{fig:redox_prediction}B). This behavior arises from diminished NADH generation due to reduced glycolytic flux, while mitochondrial oxidation of existing NADH continues. The transient imbalance between NADH production and consumption leads to a temporary elevation of the NAD$^+$ fraction before a new steady state is reached.

In contrast, under severe hypoxia (0\% O$_2$), the redox ratio exhibits a sharp decline followed by a prolonged recovery period upon restoration of normoxic conditions (Fig.~\ref{fig:redox_prediction}C). Suppression of OXPHOS under anoxic conditions limits mitochondrial NADH oxidation and NAD$^+$ regeneration, leading to an accumulation of reduced cofactors,  particularly NADH relative to NAD$^+$. This shift toward a more reduced intracellular state reflects impaired electron transport chain activity and disruption of normal mitochondrial redox balance. The delayed recovery after oxygen restoration likely reflects the time required to reestablish mitochondrial oxidative capacity and resume normal tricarboxylic acid (TCA) cycle turnover, which depends on NAD$^+$ regeneration to sustain oxidative metabolism \citep{zhang2012metabolic,Folmes2011MetabolicPlasticity}. 

Overall, the simulated redox responses capture the qualitative trends expected from redox metabolic regulation under nutrient and oxygen perturbations. Validation against experimental measurements therefore focuses on reproducing the directionality and temporal patterns of redox changes rather than achieving direct quantitative agreement. The available experimental dataset~\cite{Liu2018} was obtained from a different cell line, which may exhibit distinct metabolic capacities and baseline NAD$^+$/(NAD$^+$ + NADH) ratios. In addition, the hypoxia induction protocol used in the referenced study likely altered medium pH due to carbon dioxide depletion during prolonged nitrogen bubbling, a procedural detail that was not explicitly reported. Such pH shifts can substantially influence enzyme kinetics and redox equilibria, thereby affecting measured redox ratios. As the iPS cell line was different, consequently, model validation emphasizes consistency in dynamic redox trends rather than absolute numerical agreement.

% ===================
\subsection{Aggregate Model Prediction Performance}
\label{subsec:aggregatePrediction}

\begin{figure*}[htbp]
    \centering
    \includegraphics[width=0.72\linewidth]{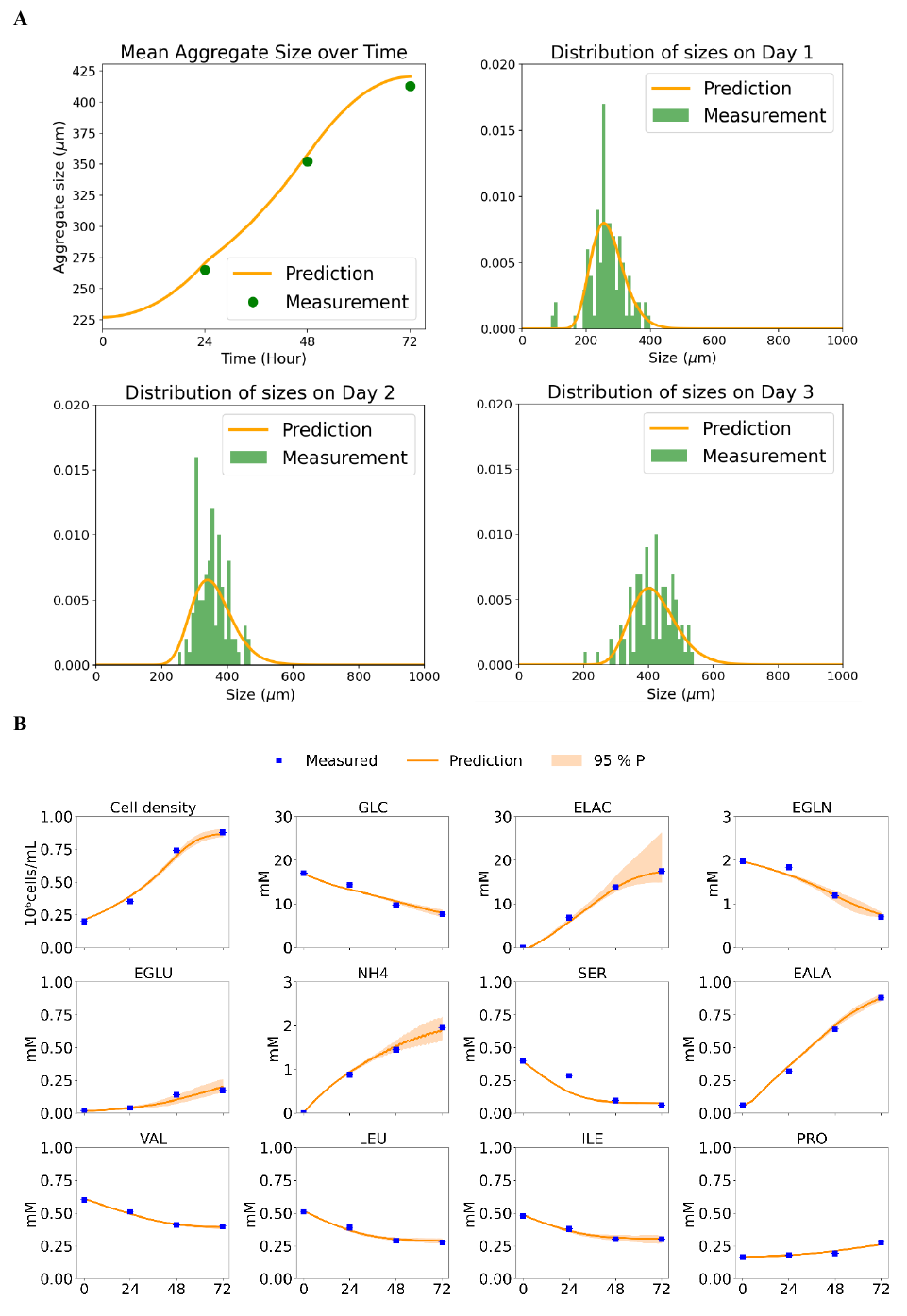}
    \caption{Aggregate growth, size distribution, and metabolic predictions compared with experimental measurements.
\textbf{(A)} Predicted and measured mean iPSC aggregate size over time, together with aggregate size distributions on Days~1--3.
Solid lines denote model predictions, data points represent experimental measurements with standard deviations.
\textbf{(B)} Predicted and measured cell density and extracellular metabolite concentrations for aggregate cultures.
 {Blue dots indicate experimental measurements from the third biological replicate, orange lines denote the mean model prediction, and shaded orange regions represent the 95\% prediction interval.}
Metabolites include extracellular glucose (GLC), extracellular lactate (ELAC), extracellular glutamine (EGLN), extracellular glutamate (EGLU), extracellular ammonia (NH4), extracellular serine (SER), extracellular alanine (EALA), extracellular valine (VAL), extracellular leucine (LEU), extracellular isoleucine (ILE), and extracellular proline (PRO). }

    \label{fig:aggregate_prediction_combined}
\end{figure*}

\textit{(1) Model prediction performance on aggregate size and distribution $\hat{\phi}(x,t)$.}
The model-predicted evolution of iPSC aggregate size over multiple culture days is shown in Fig.~\ref{fig:aggregate_prediction_combined}A. The model accurately captures both the mean aggregate size trajectory and the full size-distribution dynamics. Overall, the prediction curves align well with the experimental measurements, demonstrating strong performance in reproducing aggregate growth and distributional changes over time.

\vspace{0.1in}
\begin{sloppypar}
%\subsection*{
\textit{(2) Model prediction performance for extracellular metabolites in aggregate cultures.} 
The model’s predictions for VCD, GLC, LAC, GLN, GLU, NH$_4$, SER, ALA, VAL, LEU, ILE, and PRO are shown in Fig. \ref{fig:aggregate_prediction_combined}B.  {In the figure, blue dots denote experimental measurements from the third biological replicate of iPSC aggregate cultures, which was not used for parameter estimation and was reserved for out‑of‑sample validation. Orange lines represent the mean model prediction, and shaded orange regions indicate the 95\% prediction interval.} Overall, the model demonstrates strong predictive performance across all metabolites. Glucose and glutamine are steadily consumed, while lactate and ammonium accumulate, reflecting the characteristic metabolic behaviors of glycolysis, glutaminolysis, and nitrogen metabolism in proliferating iPSCs. Likewise, amino acids such as valine, leucine, and isoleucine show consistent depletion, indicating that the model effectively captures the dominant metabolic fluxes.  {The relatively large prediction interval observed for lactate is similar to static culture, but spatial heterogeneity in nutrient and oxygen availability may further amplify this variability.}

\end{sloppypar}

Quantitative prediction errors for all three biological replicates are summarized in Table~\ref{tab:aggregate_nmae}, with mean nMAE values ranging from 4.19\% to 4.30\%, demonstrating consistent predictive performance across replicates. Overall, the predictions for most metabolites are accurate. The relatively higher prediction error for extracellular glutamate (EGLU) likely arises from the intricate regulation of glutamine–glutamate metabolism. As a central intermediate linking carbon and nitrogen metabolic pathways, glutamate levels are highly sensitive to subtle shifts in intracellular flux distributions \cite{Yang2014}. Small deviations in pathways such as glutaminolysis, transamination, or TCA cycle anaplerosis can therefore lead to amplified changes in extracellular glutamate. Moreover, spatial heterogeneity within aggregates can further modulate local glutamine uptake and glutamate secretion, introducing additional variability that is challenging for a single-cell model to capture.

\begin{table}[htbp]
\centering
\caption{nMAE (\%) between model predictions and experimental measurements for aggregate culture variables.}
\label{tab:aggregate_nmae}

\vspace{2mm}
\setlength{\tabcolsep}{25pt}

\begin{tabular}{lcccc}
\hline
\textbf{Variable} & \textbf{Rep1} & \textbf{Rep2} & \textbf{Rep3} & \textbf{Mean} \\
\hline
Cell Density & 3.07 & 3.60 & 3.71 & 3.46 \\
GLC          & 3.03 & 5.67 & 2.95 & 3.88 \\
ELAC         & 2.18 & 1.79 & 1.55 & 1.84 \\
EGLN         & 6.29 & 3.77 & 4.22 & 4.76 \\
EGLU         & 7.59 & 9.46 & 16.14 & 11.06 \\
NH$_4$       & 2.65 & 2.65 & 2.65 & 2.65 \\
SER          & 11.09 & 10.34 & 9.43 & 10.29 \\
EALA         & 3.57 & 2.21 & 2.49 & 2.76 \\
VAL          & 1.97 & 2.04 & 1.88 & 1.96 \\
LEU          & 2.58 & 3.12 & 2.65 & 2.78 \\
ILE          & 2.06 & 2.56 & 2.13 & 2.25 \\
PRO          & 4.20 & 3.61 & 1.82 & 3.21 \\
\hline
\textbf{Mean nMAE} & \textbf{4.19} & \textbf{4.23} & \textbf{4.30} & \textbf{4.24} \\
\hline
\end{tabular}

\end{table}

\section{Discussion}

\textbf{(1) Multi-Scale Foundational Modeling and Experimental Validation.} 
This study introduces a multiscale mechanistic foundation modeling framework that characterizes iPSC manufacturing across a broad spectrum of culture configurations—from the relatively homogeneous monolayer systems commonly used in laboratory settings to the heterogeneous aggregate cultures favored for large-scale industrial production. At the core of the framework is an expanded single-cell mechanistic metabolic model that integrates extracellular culture dynamics with intracellular metabolic flux distributions and cellular redox states. To represent aggregate cultures, the framework incorporates additional multiscale modules that capture cell–cell interactions and spatial heterogeneity. A population balance model describes aggregate formation and size-distribution dynamics arising from cell aggregation and proliferation, while a reaction–diffusion model resolves nutrient transport and metabolic gradients within aggregates. Together, these modular components allow the multiscale framework to flexibly represent both monolayer and aggregate systems with diverse forms of heterogeneity.
 
To construct and validate the framework, systematic monolayer and aggregate culture experiments were performed. The resulting datasets include cell growth measurements, extracellular metabolite profiles, aggregate size dynamics, mass isotopomer distributions (MIDs) obtained from isotope tracer experiments, and non-destructive optical redox measurements. These complementary datasets provide comprehensive information for model calibration and validation across the Historic Static, Static Pyruvate, and Aggregate culture conditions. By integrating these multi-modal measurements, the framework enables quantitative evaluation of metabolic behaviors across distinct culture configurations and supports predictive modeling of iPSC metabolism under heterogeneous manufacturing environments.

\vspace{0.1in}
\textbf{(2) Monolayer Cultures and Single-cell Mechanistic Modeling.}
The relative homogeneity of monolayer cultures offers a well‑controlled setting for investigating single‑cell metabolic responses to specific extracellular perturbations. In these systems, cells experience nearly uniform concentrations of nutrients, oxygen, and metabolites, minimizing spatial heterogeneity and reducing confounding effects from transport limitations. As a result, intracellular metabolic regulation can be interpreted more directly, making monolayer cultures an ideal setting for elucidating intrinsic single‑cell metabolic behavior. These insights form the foundational building block of the broader multiscale mechanistic modeling framework.

\begin{sloppypar}
Because $^{13}$C metabolic flux analysis ($^{13}$C‑MFA) uniquely enables quantitative estimation of intracellular reaction rates, multiple isotope‑labeling experiments were employed to probe metabolic responses to environmental perturbations. Building on our previous $^{13}$C‑MFA studies of iPSC metabolism in static monolayer cultures \citep{Odenwelder2021}, a set of complementary isotopic tracers was used to interrogate distinct segments of central carbon metabolism. Specifically, [U‑$^{13}$C$_6$]glucose was applied to trace glycolytic flux and carbon entry into the TCA cycle through both pyruvate dehydrogenase (PDH) and pyruvate carboxylase (PC) pathways. 
[U‑$^{13}$C$_5$]glutamine was used to characterize glutaminolysis and anaplerotic carbon input into the TCA cycle through $\alpha$‑ketoglutarate. In addition, [1,2-$^{13}$C$_2$]glucose was used to differentiate glycolytic flux from pentose phosphate pathway (PPP) activity by generating distinct isotopomer labeling patterns in downstream metabolites such as lactate and alanine. Together, these tracers provide comprehensive coverage of carbon flux through glycolysis, the TCA cycle, anaplerotic reactions, and the pentose phosphate pathway.
\end{sloppypar}

\begin{sloppypar}
In this study, an additional tracer, [U-$^{13}$C$_3$]pyruvate, was introduced to further resolve carbon flow through pyruvate metabolism and the tricarboxylic acid (TCA) cycle. Experimental observations showed that supplementation with extracellular pyruvate significantly alters central carbon metabolism. Under Static Pyruvate conditions, glucose consumption decreased and the glutamine‑to‑glutamate conversion ratio shifted, indicating partial substrate substitution and feedback regulation of glycolytic and anaplerotic pathways. Mass isotopomer distribution (MID) analysis (Fig.~\ref{fig:mid_PYR}) confirmed incorporation of [U-$^{13}$C$_3$]pyruvate into downstream metabolites, demonstrating its participation in central carbon metabolism and constraining its partitioning between cytosolic and mitochondrial pathways. 
\end{sloppypar}

\vspace{0.1in}
\begin{sloppypar}
\textbf{(3) Extension to Heterogeneous Aggregate Cultures through Multiscale Modeling.} 
Building on the mechanistic understanding established at the single-cell level, the multi-scale modeling framework was extended to aggregate cultures by incorporating additional modules describing aggregation dynamics and intra-aggregate transport processes. In contrast to monolayer cultures, aggregate systems develop heterogeneous microenvironments due to diffusion limitations and extensive cell–cell interactions. Resulting gradients in oxygen, nutrients, and metabolic byproducts create spatial variability in cellular metabolism that cannot be captured by single‑cell models alone.
\end{sloppypar}

Experimental observations in aggregate cultures revealed coordinated metabolic adaptations driven by these heterogeneous conditions. Compared with monolayer cultures, aggregates exhibited reduced glucose consumption and lactate production, consistent with oxygen and nutrient gradients that reshape central carbon metabolism. Diffusion‑limited oxygen availability in the aggregate interior likely suppresses oxidative metabolism and alters the distribution of glycolytic flux. Reduced glutamine consumption and glutamate production further indicate a reorganization of nitrogen metabolism in response to spatial heterogeneity. Notably, ammonia production remained comparable across culture formats, suggesting compensatory interactions among glutamine uptake, transamination reactions, and downstream nitrogen processing. Collectively, these results suggest that metabolic changes in aggregates arise from redistribution of metabolic fluxes rather than uniform downregulation of metabolic activity.

Importantly, both monolayer and aggregate cultures can be represented within the unified multiscale foundational modeling framework, enabling integration of heterogeneous observations across diverse manufacturing systems and facilitating scale‑up. In monolayer cultures, homogeneous extracellular conditions allow the framework to resolve intrinsic single‑cell metabolic regulation. In aggregate cultures, additional modules describing aggregation dynamics and reaction–diffusion transport capture the heterogeneous microenvironment that shapes cellular metabolism at larger manufacturing scales. This unified representation enables the framework to flexibly describe iPSC culture behavior across manufacturing configurations, from controlled laboratory monolayer systems to heterogeneous aggregate cultures used in large‑scale bioproduction.

Collectively, this work demonstrates the value of integrating isotopic labeling, extracellular metabolite measurements, optical redox sensing, and multiscale transport modeling within a unified modeling and analytical framework. Establishing intracellular flux‑regulation mechanisms under controlled static conditions enables mechanistic interpretation of the metabolic reorganization that emerges in aggregate cultures when spatial heterogeneity is introduced. From an engineering perspective, this integrated approach provides a foundation for predictive modeling strategies that can inform monitoring, optimization, and control of scalable iPSC bioprocesses.

\vspace{0.1in}
\begin{sloppypar}
\textbf{(4) Limitations and Future Research.}
While the present framework provides a mechanistic description of intracellular metabolism, redox regulation, aggregate transport, and population dynamics, several limitations remain. The current model focuses on a defined reaction network and selected regulatory mechanisms and does not explicitly account for compartment-specific transport processes, detailed mitochondrial dynamics, or additional layers of transcriptional and signaling regulation. Incorporating these processes may further improve quantitative accuracy, particularly under dynamic or perturbed conditions.

 {In addition, the framework was calibrated and evaluated using a single iPSC cell line and specific monolayer and aggregate culture systems. 
While the underlying mechanistic structure is expected to be broadly applicable, further studies are needed to assess transferability across iPSC lines, media formulations, aggregate platforms, and bioreactor configurations. Future work will focus on systematic validation and parameter refinement using diverse experimental datasets to enhance robustness and generalizability.
}

Finally, coupling the modeling framework with real-time sensing technologies and advanced control strategies represents a promising direction toward closed-loop optimization of aggregate culture systems and the robust, scalable design of iPSC manufacturing processes.
\end{sloppypar}

\section*{Acknowledgments}

\begin{sloppypar}
This research was supported by
the National Institute of Standards and Technology (Grant Nos.~70NANB21H086, 70NANB17H002, 70NANB24H293), the National Science Foundation (Grant CAREER CMMI-2442970),
and the National Institutes of Health (NIH) through the SBIR program (Grant No.~2R44AT010840-02A1). 
The authors thank Nareg Ohannesian for his contributions to TPE redox data collection and analysis.  {The authors also thank Dustin Chang (Agilent Technologies, USA) for valuable discussions and technical support related to the experimental equipment.}
\end{sloppypar}

\section*{Appendix. Supporting information}

Supporting information associated with this article can be found in Supplementary Materials.

\section*{Data availability}

Data will be made available on request.

\bibliographystyle{unsrt}
\bibliography{cas-refs}

@article{piotrowski2022evaluation,
  title={Evaluation metrics for wind power forecasts: A comprehensive review and statistical analysis of errors},
  author={Piotrowski, Pawe{\l} and Rutyna, Inajara and Baczy{\'n}ski, Dariusz and Kopyt, Marcin},
  journal={Energies},
  volume={15},
  number={24},
  pages={9657},
  year={2022},
  publisher={MDPI}
}

@incollection{anderson2011continuous,
  title={Continuous time Markov chain models for chemical reaction networks},
  author={Anderson, David F and Kurtz, Thomas G},
  booktitle={Design and analysis of biomolecular circuits: engineering approaches to systems and synthetic biology},
  pages={3--42},
  year={2011},
  publisher={Springer}
}

@article{schwedhelm2019automated,
  title={Automated real-time monitoring of human pluripotent stem cell aggregation in stirred tank reactors},
  author={Schwedhelm, Ivo and Zdzieblo, Daniela and Appelt-Menzel, Antje and Berger, Constantin and Schmitz, Tobias and Schuldt, Bernhard and Franke, Andre and M{\"u}ller, Franz-Josef and Pless, Ole and Schwarz, Thomas and others},
  journal={Scientific reports},
  volume={9},
  number={1},
  pages={12297},
  year={2019},
  publisher={Nature Publishing Group UK London}
}

@article{mueller1984method,
  title={Method for the determination of oxygen consumption rates and diffusion coefficients in multicellular spheroids},
  author={Mueller-Klieser, W},
  journal={Biophysical journal},
  volume={46},
  number={3},
  pages={343--348},
  year={1984},
  publisher={Elsevier}
}

@article{gupta2016optimization,
  title={Optimization of agitation speed in spinner flask for microcarrier structural integrity and expansion of induced pluripotent stem cells},
  author={Gupta, Priyanka and Ismadi, Mohd-Zulhilmi and Verma, Paul J and Fouras, Andreas and Jadhav, Sameer and Bellare, Jayesh and Hourigan, Kerry},
  journal={Cytotechnology},
  volume={68},
  number={1},
  pages={45--59},
  year={2016},
  publisher={Springer}
}

@article{amit2010suspension,
  title={Suspension culture of undifferentiated human embryonic and induced pluripotent stem cells},
  author={Amit, Michal and Chebath, Judith and Margulets, Victoria and Laevsky, Ilana and Miropolsky, Yael and Shariki, Kohava and Peri, Meital and Blais, Idit and Slutsky, Guy and Revel, Michel and others},
  journal={Stem Cell Reviews and Reports},
  volume={6},
  number={2},
  pages={248--259},
  year={2010},
  publisher={Springer}
}

@article{aldous1999deterministic,
  title={Deterministic and stochastic models for coalescence (aggregation and coagulation): a review of the mean-field theory for probabilists},
  author={Aldous, David J},
  year={1999}
}

@article{xu2023high,
  title={High Nutritional Quality of Human-Induced Pluripotent Stem Cell-Generated Proteins through an Advanced Scalable Peptide Hydrogel {3D} Suspension System},
  author={Xu, Shan and Qi, Guangyan and Durrett, Timothy P and Li, Yonghui and Liu, Xuming and Bai, Jianfa and Chen, Ming-Shun and Sun, Xiuzhi and Wang, Weiqun},
  journal={Foods},
  volume={12},
  number={14},
  pages={2713},
  year={2023},
  publisher={MDPI}
}

@article{nolan2011dynamic,
  title={Dynamic model of {CHO} cell metabolism},
  author={Nolan, Ryan P and Lee, Kyongbum},
  journal={Metabolic Engineering},
  volume={13},
  number={1},
  pages={108--124},
  year={2011},
  publisher={Elsevier}
}

@article{tosaka1982analysis,
  title={Analysis of a nonlinear diffusion problem with Michaelis-Menten kinetics by an integral equation method},
  author={Tosaka, Nobuyoshi and Miyake, Shuhei},
  journal={Bulletin of Mathematical Biology},
  volume={44},
  number={6},
  pages={841--849},
  year={1982},
  publisher={Elsevier}
}

@article{ramkrishna2014population,
  title={Population balance modeling: current status and future prospects},
  author={Ramkrishna, Doraiswami and Singh, Meenesh R},
  journal={Annual Review of Chemical and Biomolecular Engineering},
  volume={5},
  pages={123--146},
  year={2014},
  publisher={Annual Reviews}
}

@article{shyh2013stem,
  title={Stem cell metabolism in tissue development and aging},
  author={Shyh-Chang, Ng and Daley, George Q and Cantley, Lewis C},
  journal={Development},
  volume={140},
  number={12},
  pages={2535--2547},
  year={2013},
  publisher={Company of Biologists}
}

@article{quinn2013quantitative,
  title={Quantitative metabolic imaging using endogenous fluorescence to detect stem cell differentiation},
  author={Quinn, Kyle P and Sridharan, Gautham V and Hayden, Rebecca S and Kaplan, David L and Lee, Kyongbum and Georgakoudi, Irene},
  journal={Scientific Reports},
  volume={3},
  number={1},
  pages={3432},
  year={2013},
  publisher={Nature Publishing Group UK London}
}

@article{takahashi2007induction,
  title={Induction of pluripotent stem cells from adult human fibroblasts by defined factors},
  author={Takahashi, Kazutoshi and Tanabe, Koji and Ohnuki, Mari and Narita, Megumi and Ichisaka, Tomoko and Tomoda, Kiichiro and Yamanaka, Shinya},
  journal={Cell},
  volume={131},
  number={5},
  pages={861--872},
  year={2007},
  publisher={Elsevier}
}

@article{yu2007induced,
  title={Induced pluripotent stem cell lines derived from human somatic cells},
  author={Yu, Junying and Vodyanik, Maxim A and Smuga-Otto, Kim and Antosiewicz-Bourget, Jessica and Frane, Jennifer L and Tian, Shulan and Nie, Jeff and Jonsdottir, Gudrun A and Ruotti, Victor and Stewart, Ron and others},
  journal={Science},
  volume={318},
  number={5858},
  pages={1917--1920},
  year={2007},
  publisher={American Association for the Advancement of Science}
}

@article{chen2011chemically,
  title={Chemically defined conditions for human {iPSC} derivation and culture},
  author={Chen, Guokai and Gulbranson, Daniel R and Hou, Zhonggang and Bolin, Jennifer M and Ruotti, Victor and Probasco, Mitchell D and Smuga-Otto, Kimberly and Howden, Sara E and Diol, Nicole R and Propson, Nicholas E and others},
  journal={Nature Methods},
  volume={8},
  number={5},
  pages={424--429},
  year={2011},
  publisher={Nature Publishing Group US New York}
}

@article{serra2012process,
  title={Process engineering of human pluripotent stem cells for clinical application},
  author={Serra, Margarida and Brito, Catarina and Correia, Cl{\'a}udia and Alves, Paula M},
  journal={Trends in Biotechnology},
  volume={30},
  number={6},
  pages={350--359},
  year={2012},
  publisher={Elsevier}
}

@article{olmer2012suspension,
  title={Suspension culture of human pluripotent stem cells in controlled, stirred bioreactors},
  author={Olmer, Ruth and Lange, Andreas and Selzer, Sebastian and Kasper, Cornelia and Haverich, Axel and Martin, Ulrich and Zweigerdt, Robert},
  journal={Tissue Engineering Part C: Methods},
  volume={18},
  number={10},
  pages={772--784},
  year={2012},
  publisher={SAGE Publications Sage CA: Los Angeles, CA}
}

@article{wu2014oxygen,
  title={Oxygen transport and stem cell aggregation in stirred-suspension bioreactor cultures},
  author={Wu, Jincheng and Rostami, Mahboubeh Rahmati and Cadavid Olaya, Diana P and Tzanakakis, Emmanuel S},
  journal={PLoS ONE},
  volume={9},
  number={7},
  pages={e102486},
  year={2014},
  publisher={Public Library of Science San Francisco, USA}
}

@article{manstein2021high,
  title={High density bioprocessing of human pluripotent stem cells by metabolic control and in silico modeling},
  author={Manstein, Felix and Ullmann, Kevin and Kropp, Christina and Halloin, Caroline and Triebert, Wiebke and Franke, Annika and Farr, Clara-Milena and Sahabian, Anais and Haase, Alexandra and Breitkreuz, Yannik and others},
  journal={Stem Cells Translational Medicine},
  volume={10},
  number={7},
  pages={1063--1080},
  year={2021},
  publisher={Oxford University Press}
}

@article{zhang2012metabolic,
  title={Metabolic regulation in pluripotent stem cells during reprogramming and self-renewal},
  author={Zhang, Jin and Nuebel, Esther and Daley, George Q and Koehler, Carla M and Teitell, Michael A},
  journal={Cell Stem Cell},
  volume={11},
  number={5},
  pages={589--595},
  year={2012},
  publisher={Elsevier}
}

@article{yang2026multiscale,
  title={Multiscale Modeling Guided Potency Assessment of mRNA-Lipid Nanoparticles},
  author={Yang, Yuling and Qiu, Yuchen and Wang, Keqi and Liu, Yifang and Sanyal, Gautam and Whitford, Paul C and Rouhanifard, Sara H and Xie, Wei},
  journal={Molecular Therapy Nucleic Acids},
  year={2026},
  publisher={Elsevier}
}

@article{kolenc2019evaluating,
  title={Evaluating cell metabolism through autofluorescence imaging of NAD (P) H and FAD},
  author={Kolenc, Olivia I and Quinn, Kyle P},
  journal={Antioxidants \& Redox Signaling},
  volume={30},
  number={6},
  pages={875--889},
  year={2019},
  publisher={SAGE Publications Sage CA: Los Angeles, CA}
}

@article{blacker2016investigating,
  title={Investigating mitochondrial redox state using NADH and NADPH autofluorescence},
  author={Blacker, Thomas S and Duchen, Michael R},
  journal={Free Radical Biology and Medicine},
  volume={100},
  pages={53--65},
  year={2016},
  publisher={Elsevier}
}

@article{nogueira2019strategies,
  title={Strategies for the expansion of human induced pluripotent stem cells as aggregates in single-use Vertical-Wheel™ bioreactors},
  author={Nogueira, Diogo ES and Rodrigues, Carlos AV and Carvalho, Marta S and Miranda, Cl{\'a}udia C and Hashimura, Yas and Jung, Sunghoon and Lee, Brian and Cabral, Joaquim MS},
  journal={Journal of biological engineering},
  volume={13},
  number={1},
  pages={74},
  year={2019},
  publisher={Springer}
}

@article{ghorbaniaghdam2014analyzing,
  title={Analyzing clonal variation of monoclonal antibody-producing {CHO} cell lines using an in silico metabolomic platform},
  author={Ghorbaniaghdam, Atefeh and Chen, Jingkui and Henry, Olivier and Jolicoeur, Mario},
  journal={PloS ONE},
  volume={9},
  number={3},
  pages={e90832},
  year={2014},
  publisher={Public Library of Science San Francisco, USA}
}

@article{Odenwelder2021,
  author    = {Odenwelder, Daniel C. and Lu, Xiaoming and Harcum, Sarah W.},
  title     = {Induced pluripotent stem cells can utilize lactate as a metabolic substrate to support proliferation},
  journal   = {Biotechnology Progress},
  volume    = {37},
  number    = {2},
  pages     = {e3090},
  year      = {2021},
  doi       = {10.1002/btpr.3090},
  publisher = {Wiley},
}

@article{CuestaGomez2023,
  author    = {Cuesta{-}Gomez, Nerea and Aguilar, Saleta and Subramaniam, Anushree and Quietmeyer, Julius and Lim, Seongjun Peter and Marks, Carla and Sowton, Andrew and Gleadall, Lewis and Hills, Robert and Stroud, Simon H. and Calvert, Hilary and Pereira, Maria},
  title     = {Suspension culture improves {iPSC} expansion and pluripotency phenotype},
  journal   = {Stem Cell Research \& Therapy},
  volume    = {14},
  number    = {1},
  pages     = {154},
  year      = {2023},
  doi       = {10.1186/s13287-023-03331-y},
  publisher = {Springer Nature},
}

@article{Wang2024,
  author    = {Wang, Keqi and Xie, Wei and Harcum, Sarah W.},
  title     = {Metabolic regulatory network kinetic modeling with multiple isotopic tracers for {iPSCs}},
  journal   = {Biotechnology and Bioengineering},
  volume    = {121},
  number    = {4},
  pages     = {1335--1353},
  year      = {2024},
  doi       = {10.1002/bit.28448},
  publisher = {Wiley},
}

@article{Rao2021,
  author    = {Rao, Yi and Fang, Yang and Chen, Sida and Wang, Zhenyu and Zhu, Qiqi and Ding, Liting and Niu, Yuxuan and Chen, Liying and Zhao, Hong and Cheng, Hailin and Wang, Xin and Hong, Xia},
  title     = {Excess exogenous pyruvate inhibits lactate dehydrogenase activity in live cells in an {MCT1}{-}dependent manner},
  journal   = {Journal of Biological Chemistry},
  volume    = {297},
  number    = {1},
  pages     = {100823},
  year      = {2021},
  doi       = {10.1016/j.jbc.2021.100823},
  publisher = {Elsevier},
}

@article{Yang2014,
  author    = {Yang, Chendong and Ko, Bakhos and Hensley, Christopher T. and Jiang, Linghao and Wasti, Ahsen T. and Kim, Jung-whan and Sudderth, Jennifer and Calvaruso, Maria Angela and Lumata, Lloyd and Mitsche, Mary and Rutter, Jared and Merritt, Matthew E. and DeBerardinis, Ralph J.},
  title     = {Glutamine oxidation maintains the TCA cycle and cell survival during impaired mitochondrial pyruvate transport},
  journal   = {Molecular Cell},
  volume    = {56},
  number    = {3},
  pages     = {414--424},
  year      = {2014},
  doi       = {10.1016/j.molcel.2014.09.025},
  publisher = {Cell Press},
}

@article{Buescher2015,
  author    = {Buescher, Joerg M. and Antoniewicz, Maciek R. and Boros, Laszlo G. and Burgess, Samuel C. and Brunengraber, Henriette and Clish, Clary B. and DeBerardinis, Ralph J. and Ferry, James G. and Frezza, Christian and Gengo, Edgar and Gottlieb, Eyal and Hiller, Karsten and Lee, Wenyun Nishino and Lorkova, Lenka and Marcucci, Elisa and Soga, Tomoyoshi and Young, J. Bryant and Zamboni, Nicola},
  title     = {A roadmap for interpreting 13{C} metabolite labeling patterns from cells},
  journal   = {Current Opinion in Biotechnology},
  volume    = {34},
  pages     = {189--201},
  year      = {2015},
  doi       = {10.1016/j.copbio.2015.02.003},
  publisher = {Elsevier},
}

@article{Vacanti2014,
  author    = {Vacanti, Nathaniel M. and Divakaruni, Ajit S. and Green, Curtis R. and Parker, Scott J. and Henry, R. Ryan and Kim, Jung-whan and Kalhan, Satish C. and Wilson, Matthew H. and MacLean, Nathan E. and Murphy, A. Matthew and Rutter, Jared and Thompson, Christopher B.},
  title     = {Regulation of substrate utilization by the mitochondrial pyruvate carrier},
  journal   = {Molecular Cell},
  volume    = {56},
  number    = {3},
  pages     = {425--435},
  year      = {2014},
  doi       = {10.1016/j.molcel.2014.09.026},
  publisher = {Cell Press},
}

@article{Liu2018,
  author    = {Liu, Zhiyi and Pouli, Dimitra and Alonzo, Carlo A. and Varone, Antonio and Karaliota, Sevasti and Quinn, Kyle P. and Münger, Karl and Karalis, Katia P. and Georgakoudi, Irene},
  title     = {Mapping metabolic changes by noninvasive, multiparametric, high-resolution imaging using endogenous contrast},
  journal   = {Science Advances},
  volume    = {4},
  number    = {3},
  pages     = {eaap9302},
  year      = {2018},
  doi       = {10.1126/sciadv.aap9302},
  publisher = {American Association for the Advancement of Science},
}

@article{Varum2011Energy,
  title={Energy metabolism in human pluripotent stem cells and their differentiated counterparts},
  author={Varum, S. and Rodrigues, A. S. and Moura, M. B. and et al.},
  journal={PLoS ONE},
  volume={6},
  number={6},
  pages={e20914},
  year={2011}
}

@article{Folmes2011MetabolicPlasticity,
  title={Metabolic plasticity in stem cell homeostasis and differentiation},
  author={Folmes, Clifford D. L. and Terzic, Andre},
  journal={Cell Stem Cell},
  volume={11},
  number={5},
  pages={596--606},
  year={2011}
}

@article{Skala2007InVivoNADH,
  title={In vivo multiphoton fluorescence lifetime imaging of protein-bound and free NADH in normal and pre-cancerous epithelia},
  author={Skala, Melissa C. and Riching, Kristin M. and Bird, Damian K. and Gendron-Fitzpatrick, Annette and Eickhoff, Jens and Eliceiri, Kevin W. and Keely, Patricia J. and Ramanujam, Nirmala},
  journal={Journal of Biomedical Optics},
  volume={12},
  number={2},
  pages={024014},
  year={2007}
}

@article{zheng2024stochastic,
  title={Stochastic biological system-of-systems modelling for {iPSC} culture},
  author={Zheng, Hua and Harcum, Sarah W and Pei, Jinxiang and Xie, Wei},
  journal={Communications Biology},
  volume={7},
  number={1},
  pages={39},
  year={2024},
  publisher={Nature Publishing Group UK London}
}

@article{si2010generation,
  title={Generation of human induced pluripotent stem cells by simple transient transfection of plasmid DNA encoding reprogramming factors},
  author={Si-Tayeb, Karim and Noto, Fallon K and Sepac, Ana and Sedlic, Filip and Bosnjak, Zeljko J and Lough, John W and Duncan, Stephen A},
  journal={BMC Developmental Biology},
  volume={10},
  number={1},
  pages={81},
  year={2010},
  publisher={Springer}
}

\newpage
\section*{Appendix. Supporting Information}

% Reset counters
\setcounter{figure}{0}
\setcounter{table}{0}
\setcounter{equation}{0}

% Redefine numbering with S prefix
\renewcommand{\thefigure}{S\arabic{figure}}
\renewcommand{\thetable}{S\arabic{table}}
\renewcommand{\theequation}{S\arabic{equation}}
\FloatBarrier
\begin{figure}[H]
    \centering
    \includegraphics[width=0.7\textwidth]{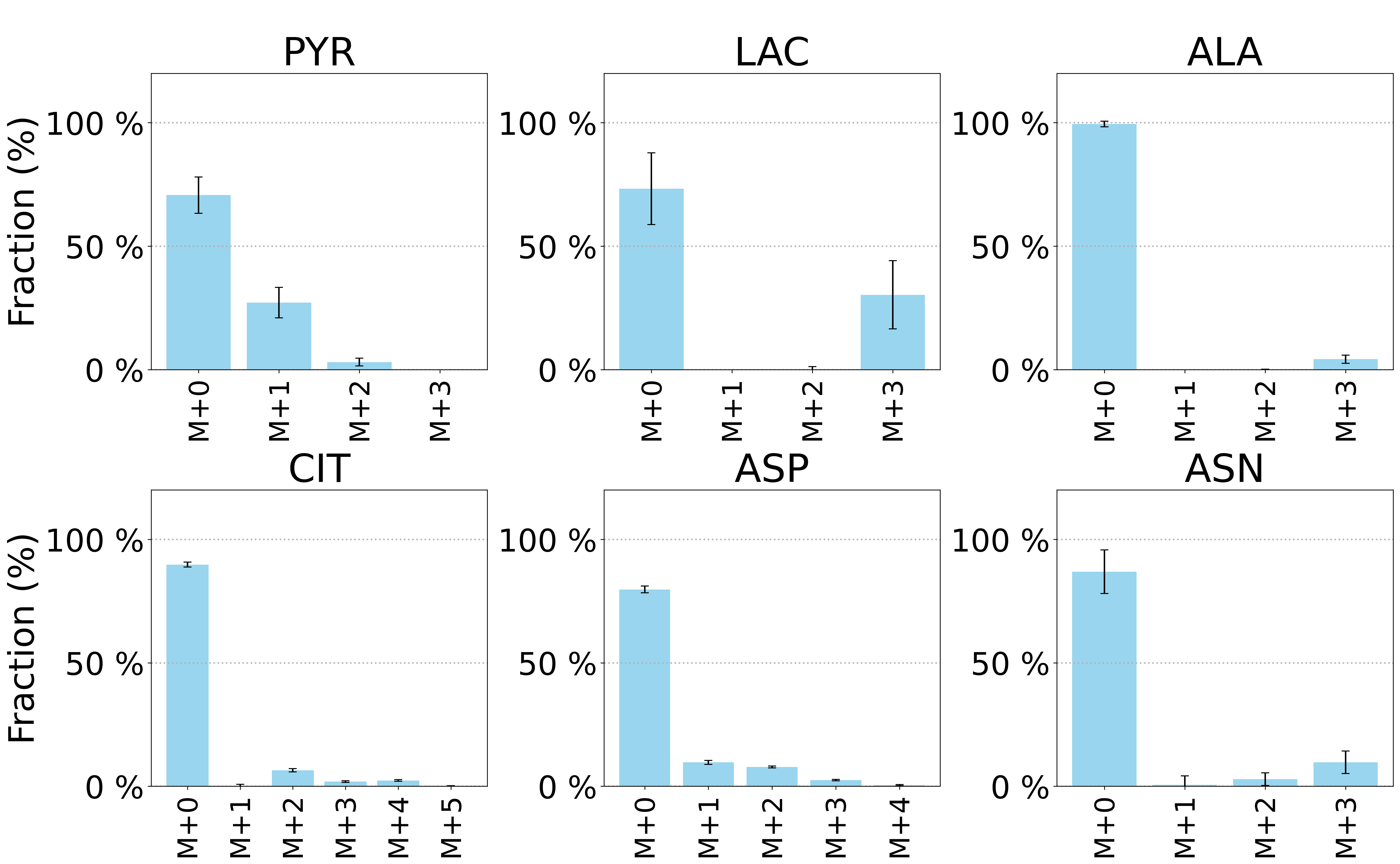}
    \caption{Intracellular Mass isotopomer distribution (MID) data for the Static Pyruvate HGHL culture condition at 48~h. MID data were corrected for natural isotopic abundance. Standard abbreviations are used for metabolites.  {Error bars represent the standard deviation across biological replicates.}}
    \label{fig:mid_hghl}
\end{figure}

\begin{figure}[H]
    \centering
    \includegraphics[width=0.7\textwidth]{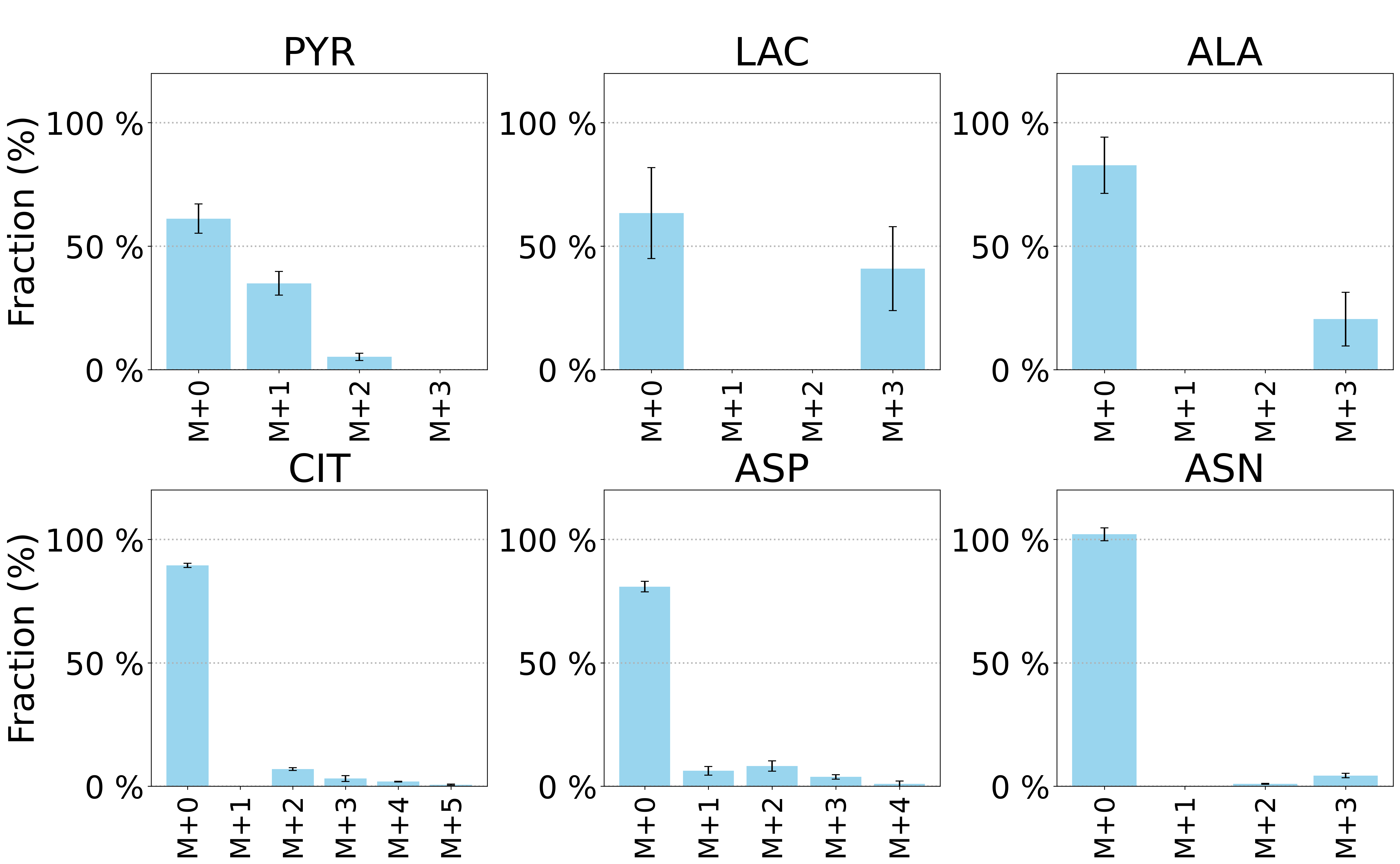}
    \caption{Intracellular Mass isotopomer distribution (MID) data for the Static Pyruvate LGHL culture condition at 48~h. MID data were corrected for natural isotopic abundance. Standard abbreviations are used for metabolites.  {Error bars represent the standard deviation across biological replicates.}}
    \label{fig:mid_lghl}
\end{figure}

\begin{figure}[H]
    \centering
    \includegraphics[width=0.7\textwidth]{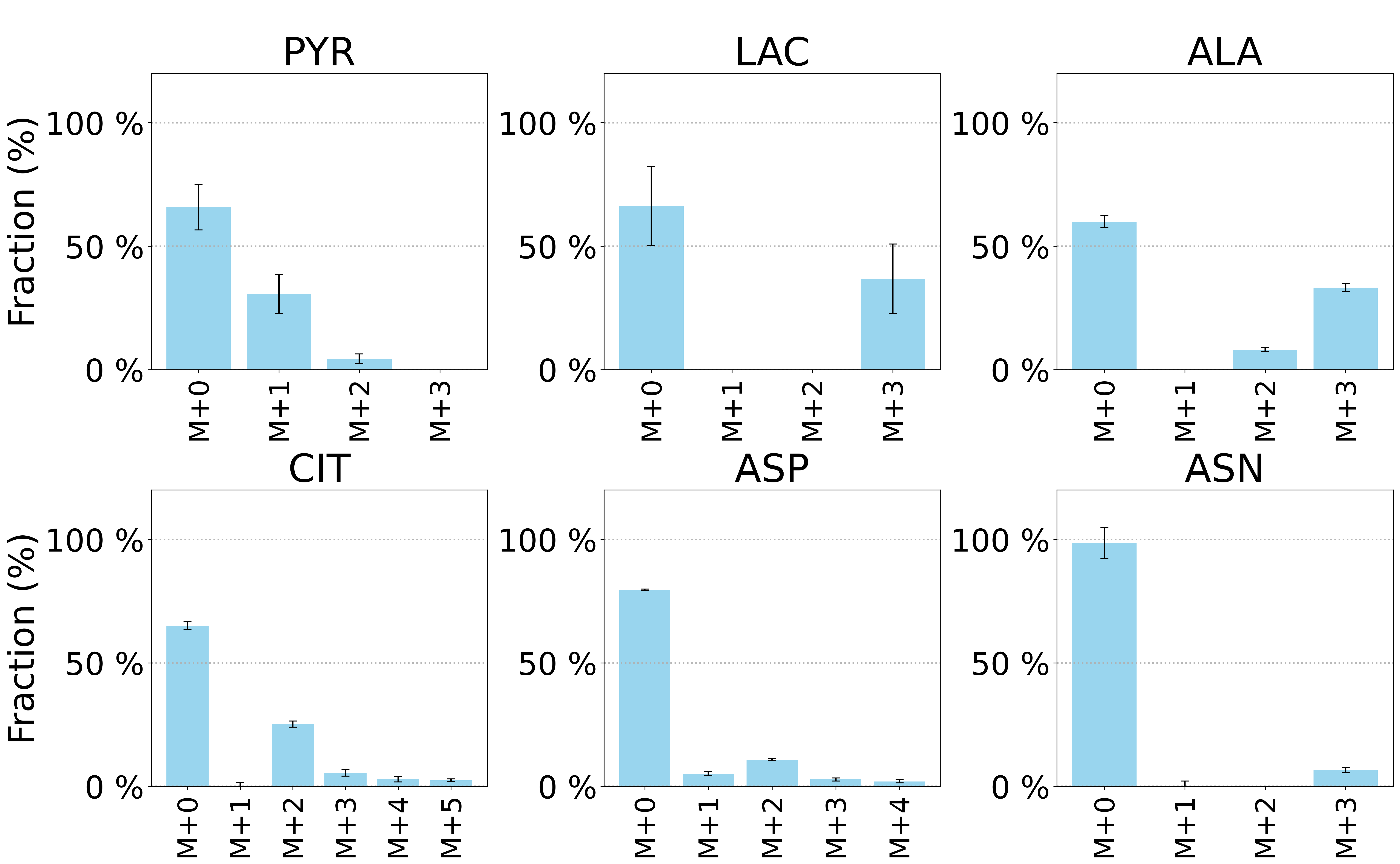}
    \caption{Intracellular Mass isotopomer distribution (MID) data for the Static Pyruvate LGLL culture condition at 48~h. MID data were corrected for natural isotopic abundance. Standard abbreviations are used for metabolites.  {Error bars represent the standard deviation across biological replicates.}}
    \label{fig:mid_lgll}
\end{figure}

\begin{figure}[H]
    \centering    \includegraphics[width=0.9\linewidth]{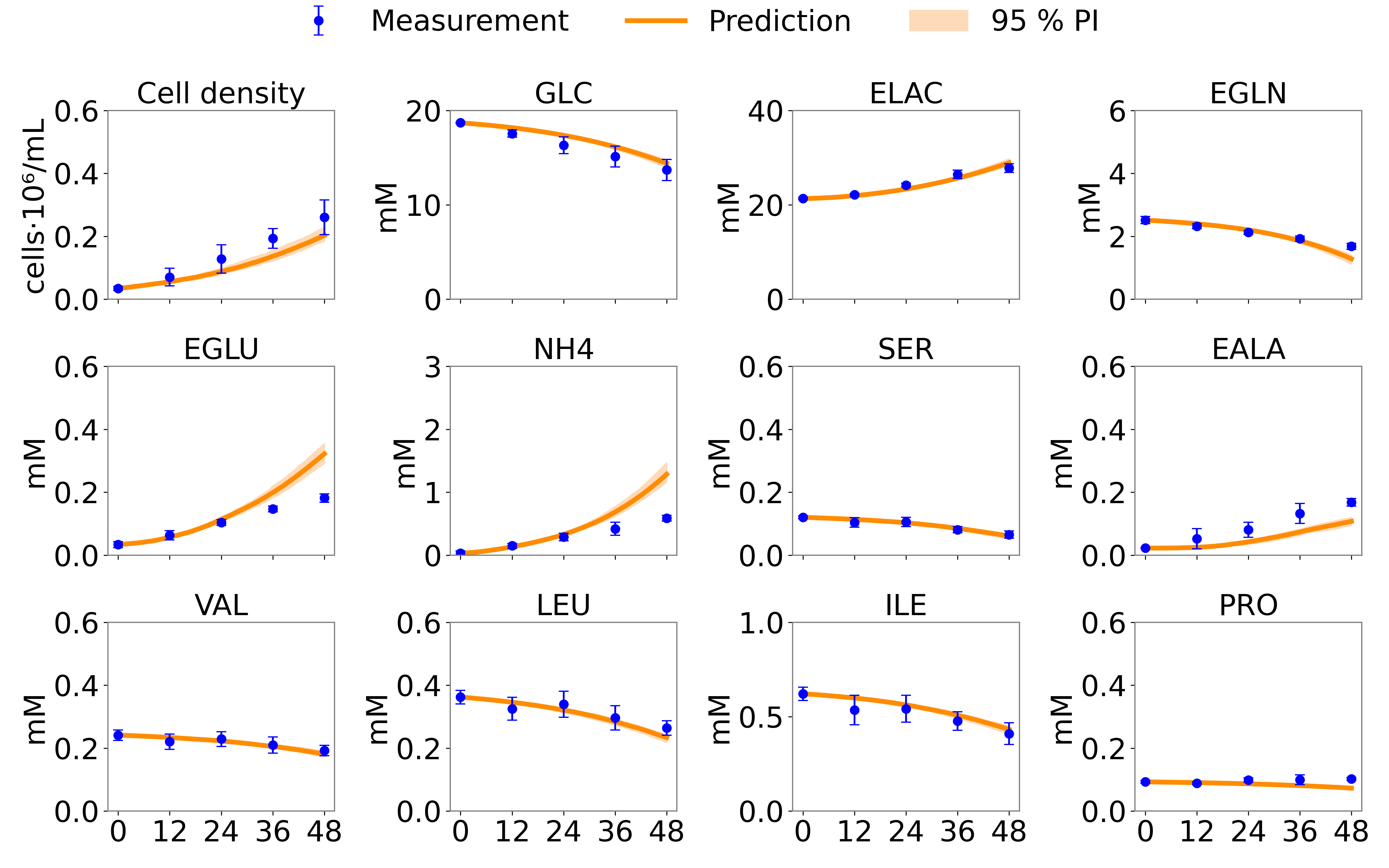}
    \caption{Dynamic model cross validation for HGHL condition left with model trained on three other Historic Static culture datasets and all Static Pyruvate culture datasets. }
    \label{fig:metaboliteprediction_no_HGLL}
\end{figure}

\begin{figure}[H]
    \centering    \includegraphics[width=0.9\linewidth]{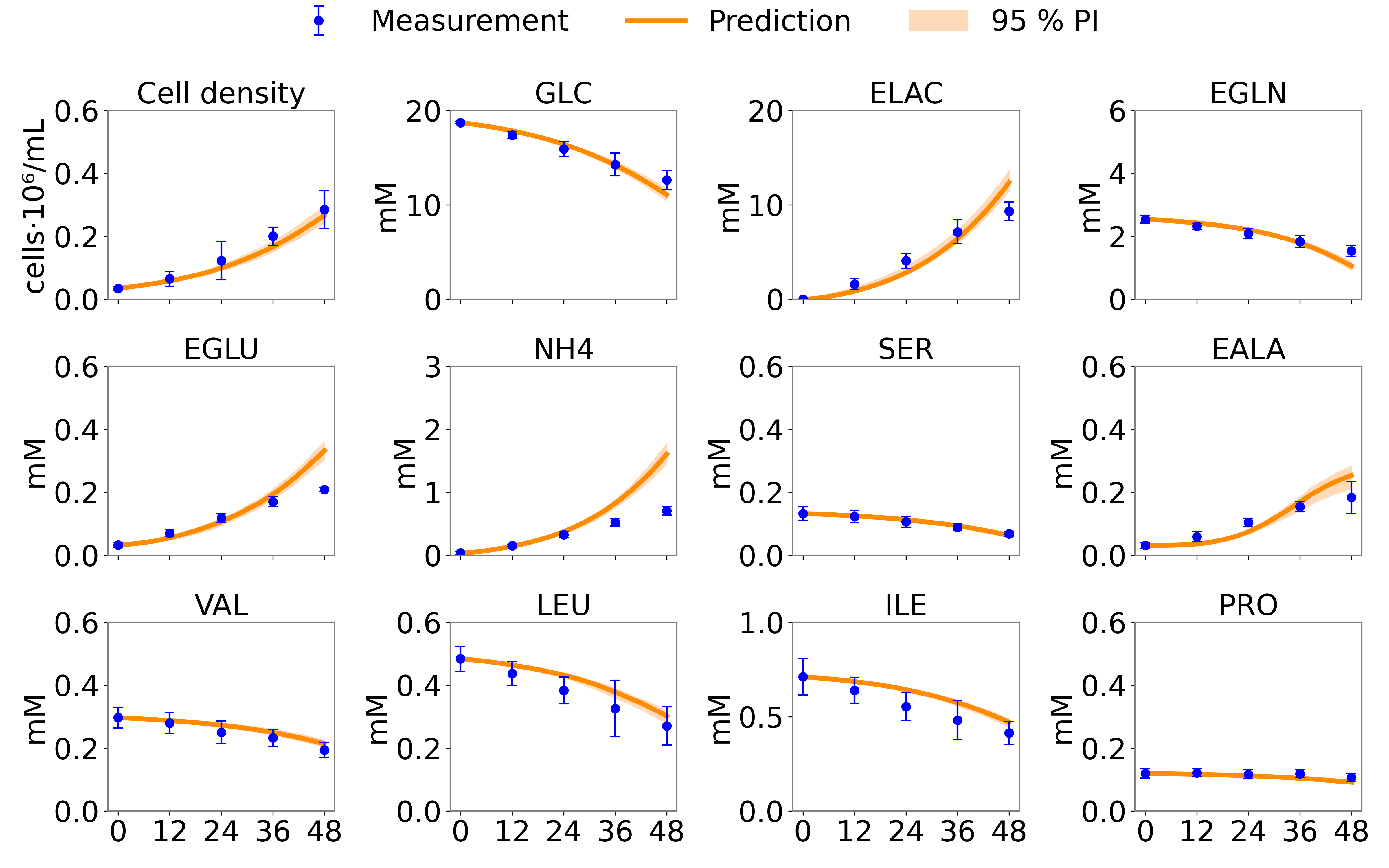}
    \caption{Dynamic model cross validation for HGLL condition left with model trained on three other Historic Static culture datasets and all Static Pyruvate culture datasets.}
    \label{fig:metaboliteprediction_no_HGLL}
\end{figure}

\begin{figure}[H]
    \centering    \includegraphics[width=0.9\linewidth]{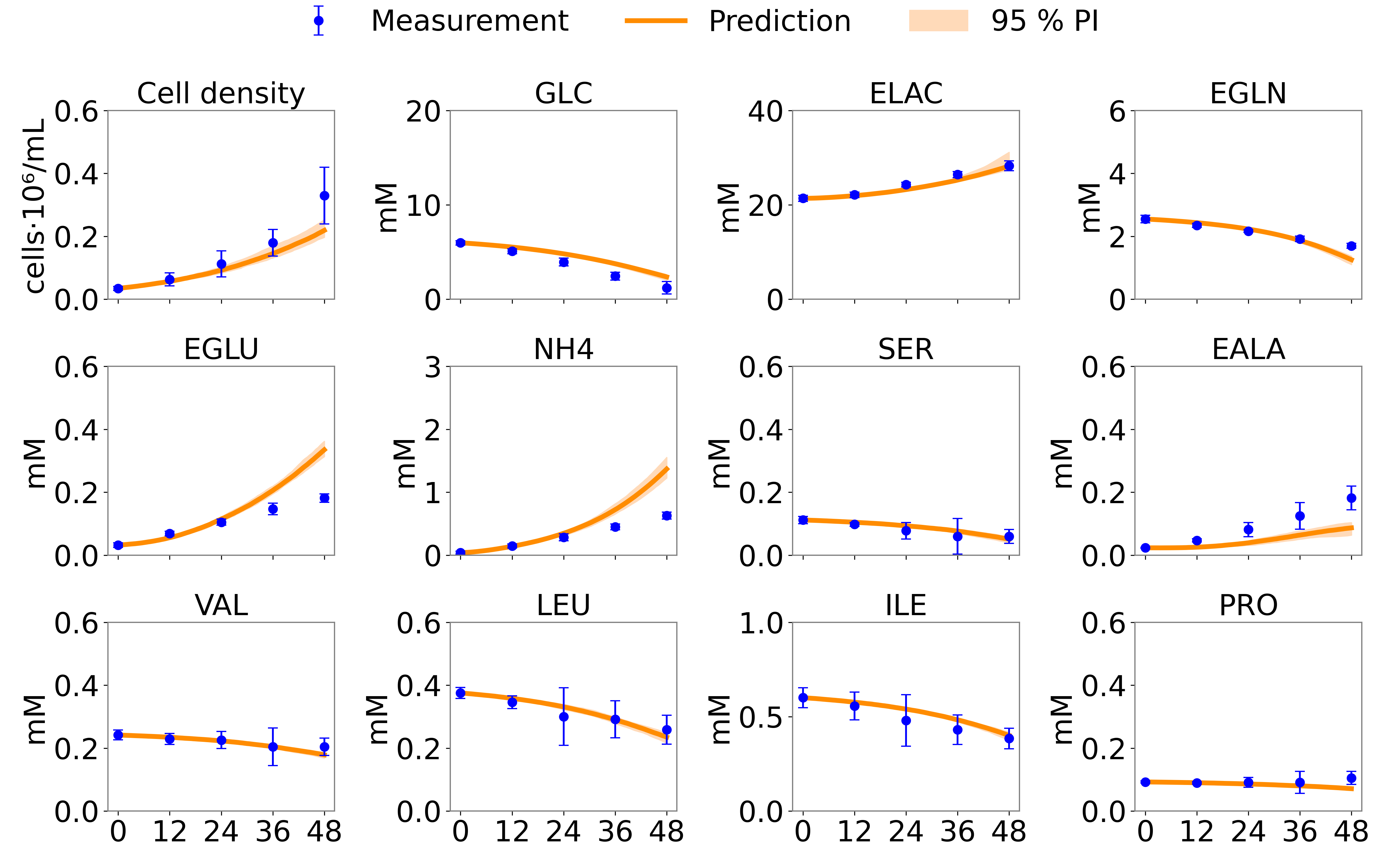}
    \caption{Dynamic model cross validation for LGHL condition left with model trained on three other Historic Static culture datasets and all Static Pyruvate culture datasets.}
    \label{fig:metaboliteprediction_no_HGLL}
\end{figure}

\begin{figure}[H]
    \centering    \includegraphics[width=0.9\linewidth]{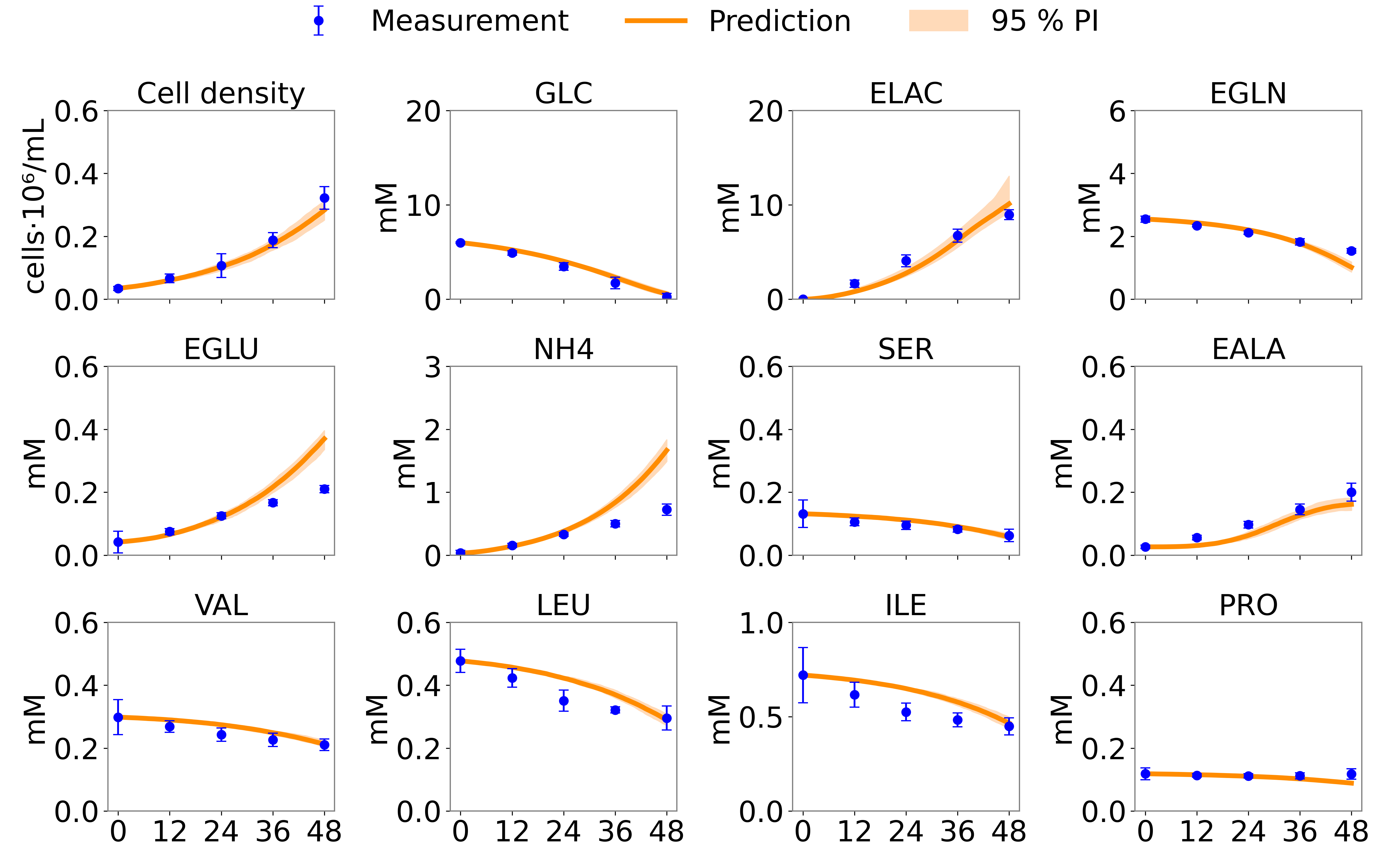}
    \caption{Dynamic model cross validation for LGLL condition left with model trained on three other Historic Static culture datasets and all Static Pyruvate culture datasets.}
    \label{fig:metaboliteprediction_no_HGLL}
\end{figure}

\begin{figure}[H]
    \centering    \includegraphics[width=0.9\linewidth]{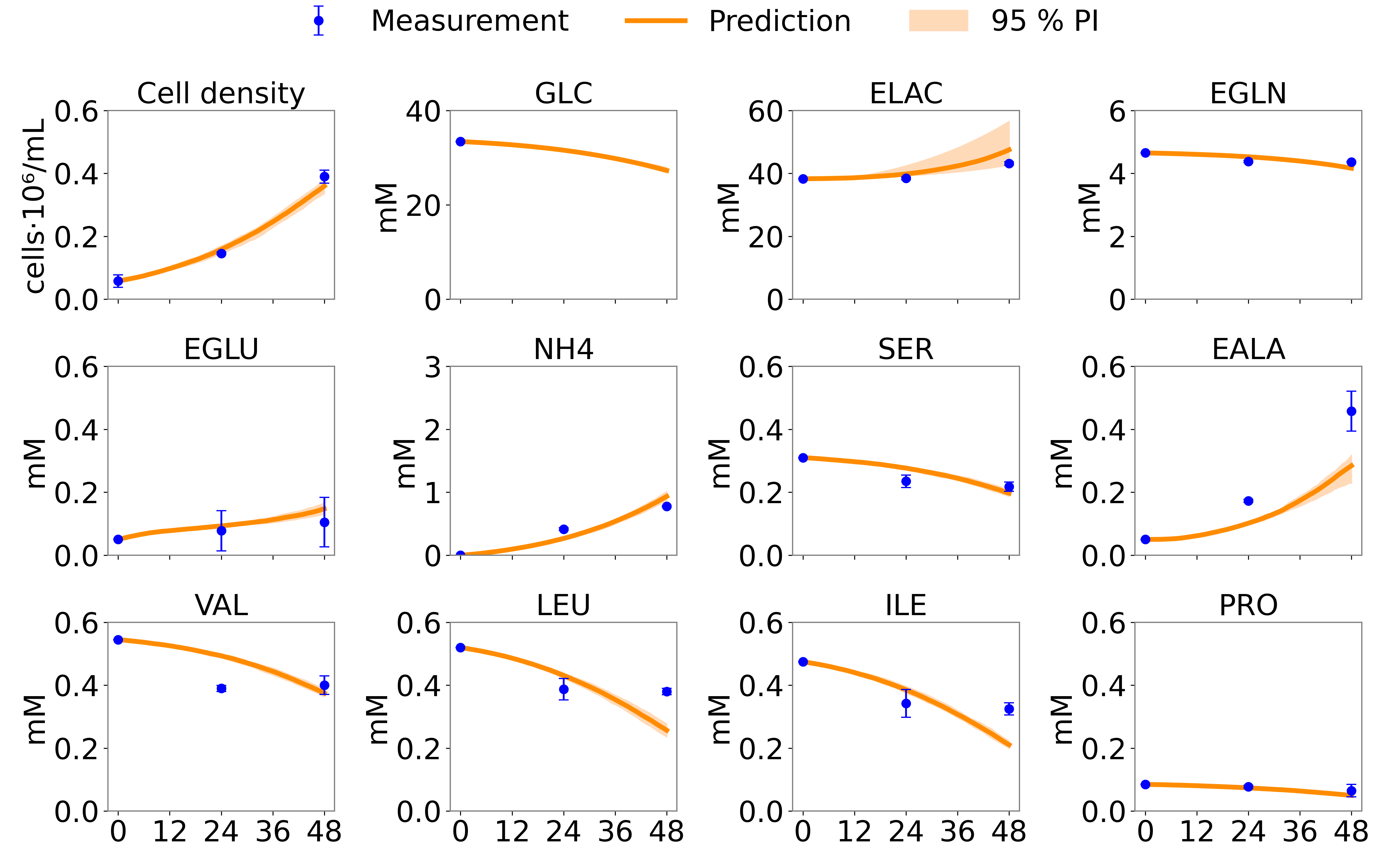}
    \caption{Dynamic model cross validation for HGHL condition left with model trained on three other Static Pyruvate Culture datasets and all Historic Static Culture datasets.}
    \label{fig:metaboliteprediction_no_HGLL}
\end{figure}

\begin{figure}[H]
    \centering    \includegraphics[width=0.9\linewidth]{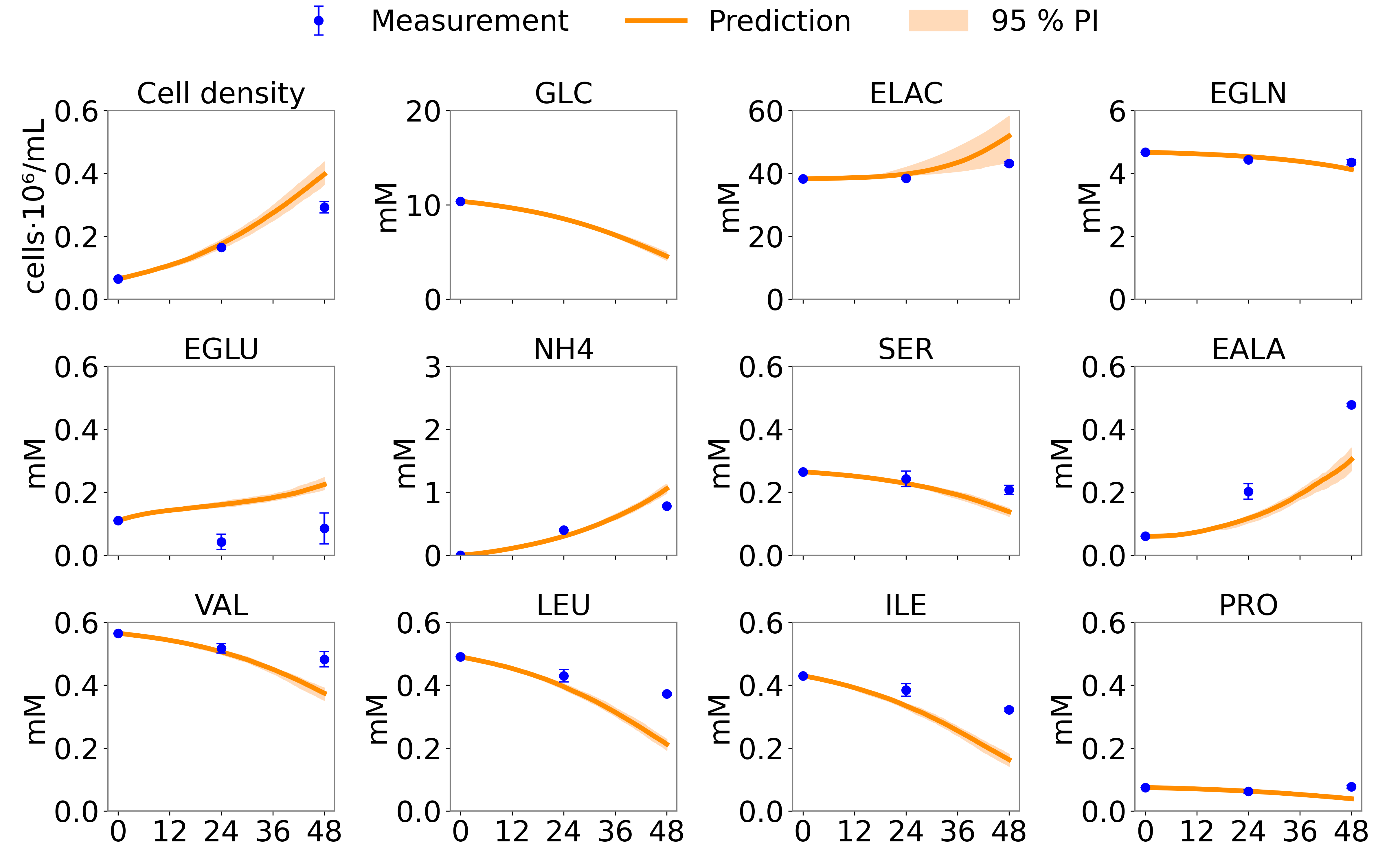}
    \caption{Dynamic model cross validation for LGHL condition left with model trained on three other Static Pyruvate Culture datasets and all Historic Static Culture datasets.}
    \label{fig:metaboliteprediction_no_HGLL}
\end{figure}

\begin{figure}[H]
    \centering    \includegraphics[width=0.9\linewidth]{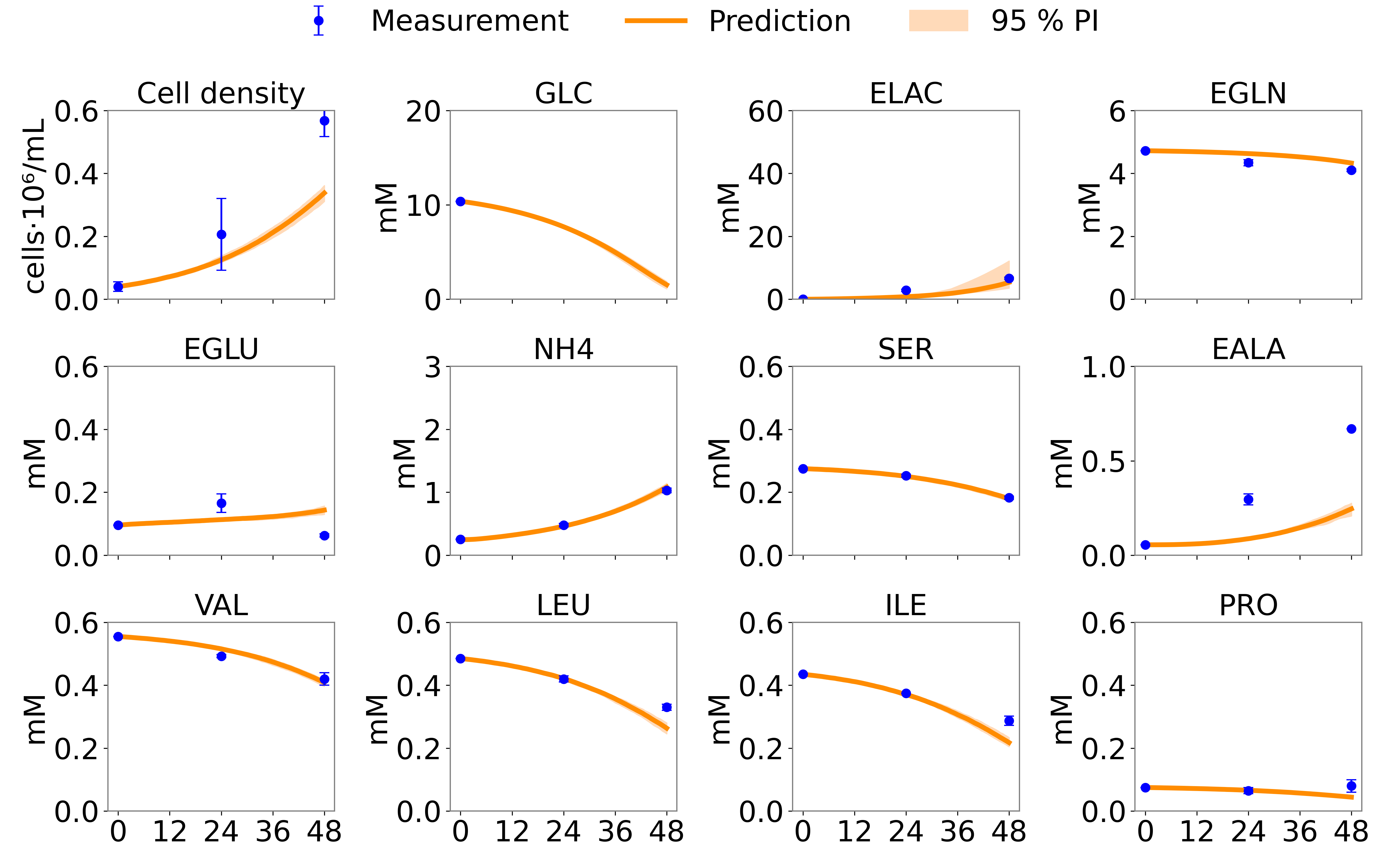}
    \caption{Dynamic model cross validation for LGLL condition left with model trained on three other Static Pyruvate Culture datasets and all Historic Static Culture datasets.}
    \label{fig:metaboliteprediction_no_HGLL}
\end{figure}

% \section*{Supplementary Tables}
\begin{table}[H]
\centering
\caption{Abbreviations for intracellular metabolites used in the MID analysis.}
\label{tab:metabolite_abbrev}
\begin{tabularx}{\textwidth}{lX lX}
\toprule
\textbf{Abbreviation} & \textbf{Metabolite name} &
\textbf{Abbreviation} & \textbf{Metabolite name} \\
\midrule
GLC & Glucose                & 3PG & 3-Phosphoglycerate \\
PEP & Phosphoenolpyruvate    & PYR & Pyruvate \\
LAC & Lactate                & ALA & Alanine \\
CIT & Citrate                & AKG & $\alpha$-Ketoglutarate \\
ACO & Aconitate              & ASP & Aspartate \\
ASN & Asparagine             & CYS & Cysteine \\
VAL & Valine                 & PHE & Phenylalanine \\
TRP & Tryptophan             & HIS & Histidine \\
\bottomrule
\end{tabularx}
\end{table}

\begin{table}[H]
\centering
\caption{\raggedright Reaction network for single-cell model including the carbon transitions.}
\label{reactiontable}
\begin{tabular}{@{}p{0.14\textwidth}p{0.86\textwidth}@{}}
\hline
\textbf{Pathway} & \textbf{Reaction} \\
\hline
\textbf{Glycolysis} &  \\
$V_1$ & $\text{GLC}{(abcdef)} + 2 \text{NAD}^+ \rightarrow  2\text{PYR}{(cba)} + \text{PYR}{(def)} + 2 \text{NADH}$ \\
$V_2$ & $\text{PYR}{(abc)} + \text{NADH} \rightleftharpoons \text{LAC}{(abc)} + \text{NAD}^+$ \\

\textbf{TCA} & \\
$V_3$ & $\text{PYR}{(abc)} + \text{NAD}^+ \rightarrow \text{ACCOA}{(bc)} + \text{NADH} + \text{CO}_2(a)$  \\
$V_4$ & $\text{PYR}{(abc)} + \text{CO}_2 (d)\rightarrow \text{OAA}{(abcd)}$\\
$V_5$ & $\text{ACCOA}{(ab)} + \text{OAA}{(cdef)} \rightarrow \text{CIT}{(fedbac)} $ \\
$V_6$ & $\text{CIT}{(abcdef)} + \text{NAD}^+ \rightarrow \text{AKG}{(abcde)} + \text{NADH} + \text{CO}_2 (f)$ \\
$V_7$ & $\text{AKG}{(abcde)} + \text{COA} + \text{NAD}^+ \rightarrow \text{SUCCOA}{(abcd)} + \text{NADH} + \text{CO}_2 (e)$ \\
$V_8$ & $\text{SUCCOA}{(abcd)} + 2/3\text{NAD}^+ \rightarrow \text{FUM}{(abcd)}  + 2/3\text{NADH}$ \\
$V_9$ & $\text{FUM}{(abcd)} \rightarrow \text{MAL}{(abcd)}$ \\
$V_{10}$ & $\text{MAL}{(abcd)} + \text{NAD}^+ \rightarrow \text{OAA}{(abcd)} + \text{NADH}$ \\
$V_{11}$ & $\text{MAL}{(abcd)} + \text{NADP}^+ \rightarrow \text{PYR}{(abc)} + \text{NADPH} + \text{CO}_2 (d)$ \\

\textbf{AA} &  \\
$V_{12}$ & $\text{GLN}{(abcde)} \rightleftharpoons \text{GLU}{(abcde)} + \text{NH}_4$ \\
$V_{130}$ & $\text{PRO}{(abcde)} + \text{NAD}^+ \rightarrow \text{GLU}{(abcde)} + \text{NADH}$ \\
$V_{14}$ & $\text{HIS}{(abcdef)} \rightarrow \text{GLU}{(abcde)} + \text{NH}_4 + \text{THF}(f)$ \\
$V_{15}$ & $\text{GLU}{(abcde)} + \text{NAD}^+ \rightarrow \text{AKG}{(abcde)} + \text{NADH} + \text{NH}_4$ \\
$V_{16}$ & $\text{GLU}{(abcde)} + \text{PYR}{(fgh)} \rightleftharpoons \text{AKG}{(abcde)} + \text{ALA}{(fgh)}$ \\
$V_{17}$ & $\text{SER}{(abc)} \rightarrow \text{PYR}{(abc)} + \text{NH}_4$ \\
$V_{18}$ & $\text{ASN}{(abcd)} \rightarrow \text{ASP}{(abcd)} + \text{NH}_4$ \\
$V_{19}$ & $\text{ASP}{(abcd)} + \text{AKG}{(efghi)} \rightleftharpoons \text{GLU}{(efghi)} + \text{OAA}{(abcd)} + \text{NH}_4$ \\
$V_{20}$ & $\text{ILE}{(abcde)} + \text{CO}_2(f) + 2\text{COA} \rightarrow \text{ACCOA}{(ab)} + \text{SUCCOA}{(fcde)} $ \\

\multicolumn{2}{@{}l}{\textbf{Oxidative Phosphorylation}} \\
$V_{21}$ & $\text{O}_2 + 2\text{NADH} \rightarrow 2\text{NAD}^+ + 2\text{H}_2\text{O}$ \\

\textbf{Transport} & \\
$V_{22}$ & $\text{EPYR}{(abc)} \rightarrow \text{PYR}{(abc)}$ \\
$V_{23}$ & $\text{ALA}{(abc)} \rightarrow \text{EALA}{(abc)}$ \\
$V_{24}$ & $\text{GLU}{(abcde)} \rightarrow \text{EGLU}{(abcde)}$ \\
$V_{25}$ & $\text{EGLN}{(abcde)} \rightarrow \text{GLN}{(abcde)}$ \\
$V_{26}$ & $\text{EASP}{(abcd)} \rightarrow \text{ASP}{(abcd)}$ \\
$V_{27}$ & $\text{LAC}{(abc)} \rightarrow \text{ELAC}{(abc)}$ \\
$V_{28}$ & $\text{O}_{2(aq)} \rightarrow \text{O}_2$ \\
\textbf{Biomass} & \\
$V_{29}$ & $0.16\text{GLC} + 0.1\text{GLN} + 0.151\text{GLU} + 0.044\text{ALA} + 0.023\text{HIS} + 0.047\text{ILE} + 0.088\text{LEU} +$ \\
   & $0.085\text{LYS} + 0.04\text{SER} + 0.033\text{TYR} + 0.05\text{VAL} + 0.039\text{THR} + 0.065\text{PRO} + 0.047\text{PHE} + 0.011\text{MET} \rightarrow X$ \\
\hline
\end{tabular}
\end{table}

\setlength{\tabcolsep}{4pt}

\begin{table*}[t]
\centering
\caption{\raggedright Flux rate models for the single-cell.}
\label{fluxtable}
\begin{tabularx}{\textwidth}{@{}c >{\raggedright\arraybackslash}X@{}}
\hline
\textbf{\#} & \textbf{Flux equation} \\
\hline

% equations 1--24 here

1  & $v(\text{HK}) = v_{\text{max HK}} \cdot \frac{\text{GLC}}{K_{m\text{GLC}} + \text{GLC}} \cdot \frac{K_{i\text{LACtoHK}}}{K_{i\text{LACtoHK}} + \text{LAC}}$ \\

2 & $\begin{aligned}[t]
v(\text{LDH}) &= v_{\text{maxf LDH}} \cdot \tfrac{\text{PYR}}{K_{m\text{PYR}} + \text{PYR}}
\cdot \tfrac{\tfrac{\text{NADH}}{\text{NAD}}}{K_{m\frac{\text{NADH}}{\text{NAD}}} + \tfrac{\text{NADH}}{\text{NAD}}} \\
&\quad - v_{\text{maxr LDH}} \cdot \tfrac{\text{LAC}}{K_{m\text{LAC}} + \text{LAC}}
\cdot \tfrac{\tfrac{\text{NAD}}{\text{NADH}}}{K_{m\frac{\text{NAD}}{\text{NADH}}} + \tfrac{\text{NAD}}{\text{NADH}}}
\end{aligned}$ \\

3  & $v(\text{PDH}) = v_{\text{max PDH}} \cdot \frac{\text{PYR}}{K_{m\text{PYR}} + \text{PYR}} \cdot
\frac{\frac{\text{NAD}}{\text{NADH}}}{K_{m\frac{\text{NAD}}{\text{NADH}}} + \frac{\text{NAD}}{\text{NADH}}}$ \\

4  & $v(\text{PC}) = v_{\text{max PC}} \cdot \frac{\text{PYR}}{K_{m\text{PYR}} + \text{PYR}}$ \\

5  & $v(\text{CS}) = v_{\text{max CS}} \cdot \frac{\text{ACCOA}}{K_{m\text{ACCOA}} + \text{ACCOA}} \cdot
\frac{\text{OAA}}{K_{m\text{OXA}} + \text{OAA}}$ \\

6  & $v(\text{CITS/ISOD}) = v_{\text{max CITS/ISOD}} \cdot \frac{\text{CIT}}{K_{m\text{CIT}} + \text{CIT}} \cdot
\frac{\frac{\text{NAD}}{\text{NADH}}}{K_{m\frac{\text{NAD}}{\text{NADH}}} + \frac{\text{NAD}}{\text{NADH}}}$ \\

7  & $v(\text{AKGDH}) = v_{\text{max AKGDH}} \cdot \frac{\text{AKG}}{K_{m\text{AKG}} + \text{AKG}} \cdot
\frac{\frac{\text{NAD}}{\text{NADH}}}{K_{m\frac{\text{NAD}}{\text{NADH}}} + \frac{\text{NAD}}{\text{NADH}}}$ \\

8  & $v(\text{SDH}) = v_{\text{max SDH}} \cdot \frac{\text{SUCCOA}}{K_{m\text{SUCCOA}} + \text{SUCCOA}} \cdot
\frac{\frac{\text{NAD}}{\text{NADH}}}{K_{m\frac{\text{NAD}}{\text{NADH}}} + \frac{\text{NAD}}{\text{NADH}}}$ \\

9  & $v(\text{FUM}) = v_{\text{max FUM}} \cdot \frac{\text{FUM}}{K_{m\text{FUM}} + \text{FUM}}$ \\

10 & $v(\text{MDH}) = v_{\text{max MDH}} \cdot \frac{\text{MAL}}{K_{m\text{MAL}} + \text{MAL}} \cdot
\frac{\frac{\text{NAD}}{\text{NADH}}}{K_{m\frac{\text{NAD}}{\text{NADH}}} + \frac{\text{NAD}}{\text{NADH}}}$ \\

11 & $v(\text{ME}) = v_{\text{max ME}} \cdot \frac{\text{MAL}}{K_{m\text{MAL}} + \text{MAL}} \cdot
\frac{\frac{\text{NADP}}{\text{NADPH}}}{K_{m\frac{\text{NADP}}{\text{NADPH}}} + \frac{\text{NADP}}{\text{NADPH}}}$ \\

12 & $\begin{aligned}[t]
v(\text{GLNS}) &= v_{\text{maxf GLNS}} \cdot 
\tfrac{\text{GLN}}{K_{m\text{GLN}} + \text{GLN}}
- v_{\text{maxr GLNS}} \cdot 
\tfrac{\text{GLU}}{K_{m\text{GLU}} + \text{GLU}}
\cdot 
\tfrac{\text{NH}_4}{K_{m\text{NH}_4} + \text{NH}_4}
\end{aligned}$ \\

13 & $v(\text{PRO}) = v_{\text{max PRO}} \cdot \frac{\text{PRO}}{K_{m\text{PRO}} + \text{PRO}} \cdot
\frac{\frac{\text{NAD}}{\text{NADH}}}{K_{m\frac{\text{NAD}}{\text{NADH}}} + \frac{\text{NAD}}{\text{NADH}}}$ \\

14 & $v(\text{HIS}) = v_{\text{max HIS}} \cdot \frac{\text{HIS}}{K_{m\text{HIS}} + \text{HIS}} \cdot
\frac{\text{AKG}}{K_{m\text{AKG}} + \text{AKG}}$ \\

15 & $\begin{aligned}[t]
v(\text{GLDH}) &= v_{\text{maxf GLDH}} \cdot 
\tfrac{\text{GLU}}{K_{m\text{GLU}} + \text{GLU}}
\cdot 
\tfrac{\tfrac{\text{NAD}}{\text{NADH}}}{K_{m\frac{\text{NAD}}{\text{NADH}}} + \tfrac{\text{NAD}}{\text{NADH}}} \\
&\quad - v_{\text{maxr GLDH}} \cdot 
\tfrac{\text{AKG}}{K_{m\text{AKG}} + \text{AKG}}
\cdot 
\tfrac{\tfrac{\text{NADH}}{\text{NAD}}}{K_{m\frac{\text{NADH}}{\text{NAD}}} + \tfrac{\text{NADH}}{\text{NAD}}}
\cdot 
\tfrac{\text{NH}_4}{K_{m\text{NH}_4} + \text{NH}_4}
\end{aligned}$ \\

16 & $\begin{aligned}[t]
v(\text{AlaTA}) &= v_{\text{maxf AlaTA}} \cdot 
\tfrac{\text{GLU}}{K_{m\text{GLU}} + \text{GLU}}
\cdot 
\tfrac{\text{PYR}}{K_{m\text{PYR}} + \text{PYR}} \\
&\quad - v_{\text{maxr AlaTA}} \cdot 
\tfrac{\text{ALA}}{K_{m\text{ALA}} + \text{ALA}}
\cdot 
\tfrac{\text{AKG}}{K_{m\text{AKG}} + \text{AKG}}
\end{aligned}$ \\

17 & $v(\text{SAL}) = v_{\text{max SAL}} \cdot \frac{\text{SER}}{K_{m\text{SER}} + \text{SER}}$ \\

18 & $v(\text{ASN}) = v_{\text{max ASN}} \cdot \frac{\text{ASN}}{K_{m\text{ASN}} + \text{ASN}}$ \\

19 & $\begin{aligned}[t]
v(\text{ASTA}) &= v_{\text{maxf ASTA}} \cdot 
\tfrac{\text{ASP}}{K_{m\text{ASP}} + \text{ASP}}
\cdot 
\tfrac{\text{NH}_4}{K_{m\text{NH}_4} + \text{NH}_4} \\
&\quad - v_{\text{maxr ASTA}} \cdot 
\tfrac{\text{AKG}}{K_{m\text{AKG}} + \text{AKG}}
\cdot 
\tfrac{\text{GLU}}{K_{m\text{GLU}} + \text{GLU}}
\cdot 
\tfrac{\text{OAA}}{K_{m\text{OXA}} + \text{OAA}}
\end{aligned}$ \\

20 & $v(\text{ILE}) = v_{\text{max ILE}} \cdot \frac{\text{ILE}}{K_{m\text{ILE}} + \text{ILE}} \cdot
\frac{\frac{\text{NAD}}{\text{NADH}}}{K_{m\frac{\text{NAD}}{\text{NADH}}} + \frac{\text{NAD}}{\text{NADH}}}$ \\

21 & $v(\text{resp}) = v_{\text{max resp}} \cdot \frac{\text{O}_2}{K_{m\text{O}_2} + \text{O}_2} \cdot
\frac{\text{NADH}}{K_{m\text{NADH}} + \text{NADH}}$ \\

22 & $v(\text{PYRT}) = v_{\text{max PYRT}} \cdot \frac{\text{EPYR}}{K_{m\text{EPYR}} + \text{EPYR}}$ \\

23 & $v(\text{ALAT}) = v_{\text{max ALAT}} \cdot \frac{\text{ALA}}{K_{m\text{ALA}} + \text{ALA}}$ \\

24 & $v(\text{GLUT}) = v_{\text{max GLUT}} \cdot \frac{\text{GLU}}{K_{m\text{GLU}} + \text{GLU}}$ \\

\hline
\end{tabularx}
\end{table*}

\setlength{\tabcolsep}{4pt}

\begin{table*}[t]
\ContinuedFloat
\centering
\caption{\raggedright Flux equations for the single-cell model (continued).}
\begin{tabularx}{\textwidth}{@{}c >{\raggedright\arraybackslash}X@{}}
\hline
\textbf{\#} & \textbf{Flux equation} \\
\hline

25 & $v(\text{GLNT}) = v_{\text{max GLNT}} \cdot \frac{\text{EGLN}}{K_{m\text{EGLN}} + \text{EGLN}}$ \\

26 & $v(\text{ASPT}) = v_{\text{max ASPT}} \cdot \frac{\text{EASP}}{K_{m\text{EASP}} + \text{EASP}}$ \\

27 & $\begin{aligned}[t]
v(\text{LACT}) &= v_{\text{maxf LACT}} \cdot 
\tfrac{\text{LAC}}{K_{m\text{LAC}} + \text{LAC}}
- v_{\text{maxr LACT}} \cdot 
\tfrac{\text{ELAC}}{K_{m\text{ELAC}} + \text{ELAC}}
\end{aligned}$ \\

28 & $v(\text{O2T}) = v_{\text{max O2T}} \cdot \frac{\text{O}_2^{aq}}{K_{m\text{O2}} + \text{O}_2^{aq}}$ \\

29 & $\begin{aligned}[t]
v(\text{growth}) &= v_{\text{max growth}} \cdot
\tfrac{\text{GLC}}{K_{m\text{GLC}} + \text{GLC}} \cdot
\tfrac{\text{GLN}}{K_{m\text{GLN}} + \text{GLN}} \cdot
\tfrac{\text{ALA}}{K_{m\text{ALA}} + \text{ALA}} \cdot
\tfrac{\text{HIS}}{K_{m\text{HIS}} + \text{HIS}} \cdot
\tfrac{\text{ILE}}{K_{m\text{ILE}} + \text{ILE}} \cdot
\tfrac{\text{LEU}}{K_{m\text{LEU}} + \text{LEU}} \cdot
\tfrac{\text{LYS}}{K_{m\text{LYS}} + \text{LYS}} \\
&\quad \cdot
\tfrac{\text{SER}}{K_{m\text{SER}} + \text{SER}} \cdot
\tfrac{\text{THR}}{K_{m\text{THR}} + \text{THR}} \cdot
\tfrac{\text{TYR}}{K_{m\text{TYR}} + \text{TYR}} \cdot
\tfrac{\text{VAL}}{K_{m\text{VAL}} + \text{VAL}} \cdot
\tfrac{\text{MET}}{K_{m\text{MET}} + \text{MET}} \cdot
\tfrac{\text{PHE}}{K_{m\text{PHE}} + \text{PHE}} \cdot
\tfrac{\text{PRO}}{K_{m\text{PRO}} + \text{PRO}}
\end{aligned}$ \\

\hline
\end{tabularx}
\end{table*}

\begin{table*}[ht]
\centering
\footnotesize
\caption{ {Bootstrap uncertainty summary for the cell metabolic kinetic model parameters.}}
\label{tab:parameters_ci}
\begin{tabularx}{\textwidth}{l r l @{\hspace{10em}} l r l}
\toprule
\textbf{Parameter} & \textbf{Median} & \textbf{95\% CI} &
\textbf{Parameter} & \textbf{Median} & \textbf{95\% CI} \\
\midrule
$v_{\text{maxHK}}$ & 1.714 & [1.616, 1.776] &
$K_{i\text{LACtoHK}}$ & 32.646 & [28.679, 34.497] \\

$K_{m\text{GLC}}$ & 1.469 & [1.353, 1.526] &
$v_{\text{maxfLDH}}$ & 3.300 & [3.155, 3.345] \\

$v_{\text{maxrLDH}}$ & 0.100 & [0.096, 0.104] &
$K_{m\text{PYR}}$ & 0.206 & [0.197, 0.229] \\

$K_{m\text{LAC}}$ & 0.041 & [0.039, 0.046] &
$K_{m\text{NADHtoNAD}}$ & 0.0036 & [0.0034, 0.0038] \\

$K_{m\text{NADtoNADH}}$ & 0.514 & [0.500, 0.526] &
$v_{\text{maxPDH}}$ & 0.227 & [0.209, 0.237] \\

$v_{\text{maxPC}}$ & 0.058 & [0.056, 0.063] &
$v_{\text{maxCS}}$ & 0.441 & [0.413, 0.471] \\

$K_{m\text{AcCoA}}$ & 0.091 & [0.084, 0.097] &
$K_{m\text{OAA}}$ & 0.082 & [0.078, 0.090] \\

$v_{\text{maxCITSISOD}}$ & 1.277 & [1.233, 1.398] &
$K_{m\text{CIT}}$ & 0.383 & [0.368, 0.404] \\

$v_{\text{maxAKGDH}}$ & 2.888 & [2.731, 3.061] &
$K_{m\text{AKG}}$ & 2.920 & [2.822, 3.000] \\

$v_{\text{maxSDH}}$ & 0.306 & [0.293, 0.317] &
$K_{m\text{SUCCOA}}$ & 0.161 & [0.156, 0.175] \\

$K_{m\text{FUM}}$ & 0.168 & [0.162, 0.176] &
$v_{\text{maxFUM}}$ & 0.307 & [0.294, 0.337] \\

$v_{\text{maxMDH}}$ & 1.449 & [1.383, 1.558] &
$K_{m\text{MAL}}$ & 0.103 & [0.097, 0.108] \\

$K_{m\text{NADPtoNADPH}}$ & $9.74\times10^{-4}$ & [$9.22\times10^{-4}$, $1.00\times10^{-3}$] &
$v_{\text{maxME}}$ & 0.508 & [0.478, 0.534] \\

$v_{\text{maxfGLNS}}$ & 0.579 & [0.566, 0.607] &
$v_{\text{maxrGLNS}}$ & 20.583 & [20.282, 21.571] \\

$K_{m\text{GLN}}$ & 0.266 & [0.260, 0.279] &
$K_{m\text{GLU}}$ & 0.285 & [0.270, 0.311] \\

$K_{m\text{NH4}}$ & 1.711 & [1.635, 1.751] &
$v_{\text{maxPRO}}$ & 0.00191 & [0.00187, 0.00200] \\

$v_{\text{maxPROr}}$ & 0.00187 & [0.00183, 0.00200] &
$K_{m\text{PRO}}$ & 0.101 & [0.092, 0.109] \\

$K_{i\text{O2}}$ & $4.0\times10^{-6}$ & &
$v_{\text{maxHIS}}$ & 0.031 & [0.029, 0.033] \\

$K_{m\text{HIS}}$ & 0.076 & [0.073, 0.080] &
$v_{\text{maxfGLDH}}$ & 0.247 & [0.238, 0.260] \\

$v_{\text{maxrGLDH}}$ & 0.095 & [0.089, 0.098] &
$v_{\text{maxfAlaTA}}$ & 0.822 & [0.800, 0.854] \\

$v_{\text{maxrAlaTA}}$ & 0.103 & [0.096, 0.114] &
$K_{m\text{ALA}}$ & 0.204 & [0.190, 0.223] \\

$v_{\text{maxSAL}}$ & 0.012 & [0.011, 0.014] &
$K_{m\text{SER}}$ & 0.014 & [0.013, 0.015] \\

$v_{\text{maxASN}}$ & 0.020 & [0.019, 0.022] &
$K_{m\text{ASN}}$ & 0.050 & [0.048, 0.052] \\

$v_{\text{maxfASTA}}$ & 2.880 & [2.770, 3.055] &
$v_{\text{maxrASTA}}$ & 0.029 & [0.028, 0.029] \\

$K_{m\text{ASP}}$ & 76.618 & [75.791, 77.828] &
$v_{\text{maxILE}}$ & 0.096 & [0.091, 0.104] \\

$K_{m\text{ILE}}$ & 0.569 & [0.526, 0.602] &
$v_{\text{maxLEU}}$ & 0.052 & [0.049, 0.054] \\

$K_{m\text{LEU}}$ & 0.567 & [0.521, 0.598] &
$v_{\text{maxTHR}}$ & 0.026 & [0.023, 0.027] \\

$K_{m\text{THR}}$ & 0.543 & [0.504, 0.561] &
$v_{\text{maxTRP}}$ & 0.105 & [0.097, 0.110] \\

$K_{m\text{TRP}}$ & 0.518 & [0.505, 0.550] &
$v_{\text{maxLYS}}$ & 0.050 & [0.047, 0.053] \\

$K_{m\text{LYS}}$ & 0.533 & [0.500, 0.579] &
$v_{\text{maxVAL}}$ & 0.050 & [0.047, 0.055] \\

$K_{m\text{VAL}}$ & 0.547 & [0.538, 0.575] &
$v_{\text{maxMET}}$ & 0.099 & [0.095, 0.106] \\

$K_{m\text{MET}}$ & 0.577 & [0.550, 0.586] &
$v_{\text{maxPHE}}$ & 0.049 & [0.047, 0.051] \\

$K_{m\text{PHE}}$ & 0.540 & [0.526, 0.554] &
$v_{\text{maxTYR}}$ & 0.050 & [0.046, 0.052] \\

$K_{m\text{TYR}}$ & 0.549 & [0.511, 0.589] &
$v_{\text{maxresp}}$ & 7.077 & [6.749, 7.423] \\

$K_{m\text{O2}}$ & $4.0\times10^{-6}$ &  &
$K_{m\text{NADH}}$ & $4.0\times10^{-6}$ &  \\

$v_{\text{maxPYRT}}$ & 0.178 & [0.171, 0.181] &
$K_{m\text{EPYR}}$ & 0.181 & [0.175, 0.192] \\

$v_{\text{maxAlaT}}$ & 0.347 & [0.341, 0.378] &
$v_{\text{maxGluT}}$ & 0.249 & [0.237, 0.258] \\

$v_{\text{maxGlnT}}$ & 0.363 & [0.343, 0.377] &
$v_{\text{maxASPT}}$ & 0.414 & [0.403, 0.422] \\

$K_{m\text{EGLN}}$ & 1.027 & [0.934, 1.066] &
$K_{m\text{EASP}}$ & 12.543 & [11.724, 13.305] \\

$v_{\text{maxfLACT}}$ & 2.960 & [2.890, 3.114] &
$v_{\text{maxrLACT}}$ & 0.580 & [0.553, 0.651] \\

$K_{m\text{ELAC}}$ & 0.610 & [0.555, 0.642] &
$v_{\text{maxgrowth}}$ & 0.102 & [0.098, 0.105] \\
\bottomrule
\end{tabularx}
\begin{tablenotes}
\footnotesize
\item Note: $K_{i\text{O2}}$, $K_{m\text{O2}}$, and $K_{\text{mNADH}}$ were fixed during model calibration due to poor identifiability resulting from the lack of intracellular oxygen and NADH measurements, and were therefore excluded from the bootstrap uncertainty analysis.
\end{tablenotes}
\end{table*}

\begin{table*}[t]
\centering
\footnotesize
\caption{Diffusion coefficients for extracellular metabolites at 37$^{\circ}$C ($10^{-9}$~m$^2$/s).}
\label{tab:diffusion_coeffs}
\begin{tabularx}{\textwidth}{l r @{\hspace{19em}} l r}
\toprule
\textbf{Metabolite} & \textbf{$D_i^{a}$} &
\textbf{Metabolite} & \textbf{$D_i^{a}$} \\
\midrule
Pyruvate    & 1.12  & Glucose     & 0.60 \\
Alanine     & 0.91  & Glutamine   & 0.76 \\
Aspartate   & 0.741 & Glutamate   & 0.708 \\
Valine      & 0.83  & Lactate     & 1.033 \\
Serine      & 0.891 & Ammonia     & 1.86 \\
O$_2$       & 1.163 & Lysine      & 0.626\\
Asparagine  & 0.83  & Histidine   & 0.73 \\
Isoleucine  & 0.641 & Leucine     & 0.73 \\
Tyrosine    & 0.30 & & \\
\bottomrule
\end{tabularx}
\end{table*}

\newpage
\begin{table*}[t]
\centering
\footnotesize
\setlength{\tabcolsep}{4pt}
\caption{Evaluation of metabolites for inclusion in the single cell model for species that are significantly labeled by [U-$^{13}$C$_3$] pyruvate or exhibit differential labeling between cultures with and without elevated lactate.}
\label{tab:metabolite_accumulation_ranking}

\begin{tabularx}{\textwidth}{
    l
    p{2cm}
    p{2.2cm}
    p{2cm}
    >{\raggedright\arraybackslash}X
}
\toprule
\textbf{Metabolite} &
\textbf{Significant Accumulation ($p<0.05$)} &
\textbf{Different between High/Low Lac?} &
\textbf{Considered Accumulation in Model?} &
\textbf{Notes} \\
\midrule

3PG & Yes & No & No & Low peak intensity \\
ACO & Yes & No & No & Low peak intensity \\
AKG & Yes & No & Yes & \\
ALA & Yes & Yes & Yes & \\
ASN & Yes & No & Yes &  \\
ASP & Yes & No & Yes &  \\
CIT & Yes & No & Yes &  \\
CYS & No & Yes & No & No concentration measurement available \\
FUM & No & Yes & Yes & \\
GLC & Yes & No & No & Low peak intensity \\
GLN & No & No & No &  \\
GLU & No & No & No &  \\
GLY & No & No & No & \\
HIS & Yes & No & No & Essential amino acid; low peak intensity \\
ILE & No & No & No & Essential amino acid \\
LAC & Yes & Yes & Yes &  \\
LYS & No & No & No & Essential amino acid \\
MAL & No & No & Yes & \\
MET & No & No & No & Essential amino acid \\
PEP & Yes & No & No & Low peak intensity \\
PHE & No & No & No & Essential amino acid \\
PRO & No & No & No & \\
PYR & Yes & Yes & Yes & \\
SER & No & No & No &  \\
SUC & No & Yes & Yes & \\
THR & No & No & No & Essential amino acid \\
TRP & Yes & Yes & No & Essential amino acid \\
TYR & No & No & No & \\
VAL & No & No & No & Essential amino acid \\

\bottomrule
\end{tabularx}
\end{table*}

\end{document}